\documentclass[reprint,
unsortedaddress,
 amsmath,amssymb,
 aps,
 prl,
]{revtex4-2}
\usepackage{soul}
\usepackage{graphicx}% Include figure files
\usepackage{dcolumn}% Align table columns on decimal point
\usepackage{bm}% bold math
\usepackage[colorlinks=true,citecolor=blue,linkcolor=blue,urlcolor=blue, linktocpage=true,breaklinks=true,pagebackref=false]{hyperref}
\usepackage{siunitx}
\graphicspath{{PRL_images}}
\newcommand{\ee}{{\rm e}}

\begin{document}

\preprint{APS/123-QED}

\title{From Local to Global Chaos in a Parametric Micromechanical Resonator}% Force line breaks with \\

\author{Amit Shachar}
\affiliation{
Ben-Gurion University of the Negev, Beer-Sheva 84105, Israel
}
\author{Gabriel Margiani}
\affiliation{
Laboratory for Solid State Physics, ETH Zürich, CH-8093 Zürich, Switzerland
}
\author{Alexander Eichler}
\affiliation{
Laboratory for Solid State Physics, ETH Zürich, CH-8093 Zürich, Switzerland
}
\author{Oriel Shoshani}
\affiliation{
Ben-Gurion University of the Negev, Beer-Sheva 84105, Israel
}

\date{\today}

\begin{abstract}
We establish and experimentally validate a predictive framework for engineering chaos in parametric resonators. A rotating-frame description reveals a tunable double-well structure. An energy-balance condition yields a closed-form Melnikov threshold for global chaos. Below it, period-doubling cascades generate chaos confined to either well; above it, a merging crisis unites the local attractors. Experiments in a micromechanical resonator confirm the predictions linking local instabilities to global transport and opening a route to tailored chaotic dynamics in micromechanical resonators.
\end{abstract}

\maketitle

Chaos is often pictured as disorder: a trajectory that wanders unpredictably through all accessible states. Yet chaotic motion has an internal order. When chaos emerges in a driven nonlinear oscillator, the chaotic motion can remain confined to a region around a single state, bounded by an unstable orbit. The set of states visited in such a region is called a chaotic attractor. Several such regions can coexist, each with its own chaotic attractor. As a control parameter is tuned, the attractors can reach the boundary and merge into a single attractor that spans several states. How such chaos is organized, and how it fuses, is a fundamental question in nonlinear dynamics~\cite{meiss2015thirty,hiraiwa2024designing}.

A well-studied dynamical system that can shed light on this question is the Duffing-Holmes oscillator: a damped particle in a symmetric double-well potential, driven periodically. Chaos was first found to be confined to a single well, reached via a cascade of period-doubling bifurcations~\cite{holmes1979nonlinear,moon1985fractal}. Because the two wells are related by symmetry, each well can host a chaotic attractor. When the drive amplitude is increased, the two attractors reach the separatrix, the unstable orbit between the wells, and merge in an attractor-merging crisis. Beyond the crisis, the motion switches intermittently between the two halves of the merged attractor, with a mean residence time that follows a power law in the distance to the threshold~\cite{grebogi1983crises,grebogi1987critical,rempel2005intermittency}. Micromechanical and nanomechanical resonators now allow such dynamics to be tested experimentally with high fidelity and controllability~\cite{houri2020generic,defoort2021dynamical,wu2024true}.

Chaotic attractors, the merging crisis, and cross-well chaos have each been studied in a range of systems. However, quantifying the connection between these phenomena with analytical theory and testing the connection in an experiment has remained difficult. The Melnikov function, the standard analytical tool for cross-well chaos, treats damping and drive as small perturbations of a conservative system, so it is unclear whether it applies to real, dissipative devices. In particular, it remains open how to predict the drive amplitude at which single-well chaotic attractors merge into cross-well chaos, and how the motion beyond this threshold depends on the distance from the threshold. A suitable experiment must feature a high degree of control, sufficiently small damping and low noise, and precise readout of the dynamics over long timescales. 

In this Letter, we close this gap with a demonstration based on a parametrically driven micromechanical resonator. In a frame rotating at the drive frequency, the slow dynamics of the resonator are similar to those of a damped particle in a double-well quasi-potential. A weak, near-resonant direct drive perturbs this motion, giving rise to an effective, damped and driven Duffing-Holmes-like oscillator in the rotating frame. We calculate the period-doubling instability of a periodic orbit in one well, which starts a cascade toward a chaotic attractor confined to that well. We then apply the Melnikov function to the separatrix of the rotating-frame double well and obtain, in closed form, the direct-drive amplitude at which cross-well chaos sets in. This amplitude agrees closely with the drive at which the two chaotic attractors merge, within about 1\% in numerical simulations and about 3\% in the experiment, even though the resonator is dissipative. Beyond the merging, the mean residence time in each well follows a power law in the distance from the threshold, with an exponent that we calculate from the Floquet multipliers of the saddle periodic orbit and confirm experimentally. These results provide a quantitative, experimentally verified route to tailoring the extent of chaos and the time scales of switching.

\begin{figure}[t]
    \centering
     % (a) Experimental setup
   \includegraphics[
    width=0.95\columnwidth,
    trim={9.80cm 13.10cm 11.25cm 1.15cm},
    clip
]{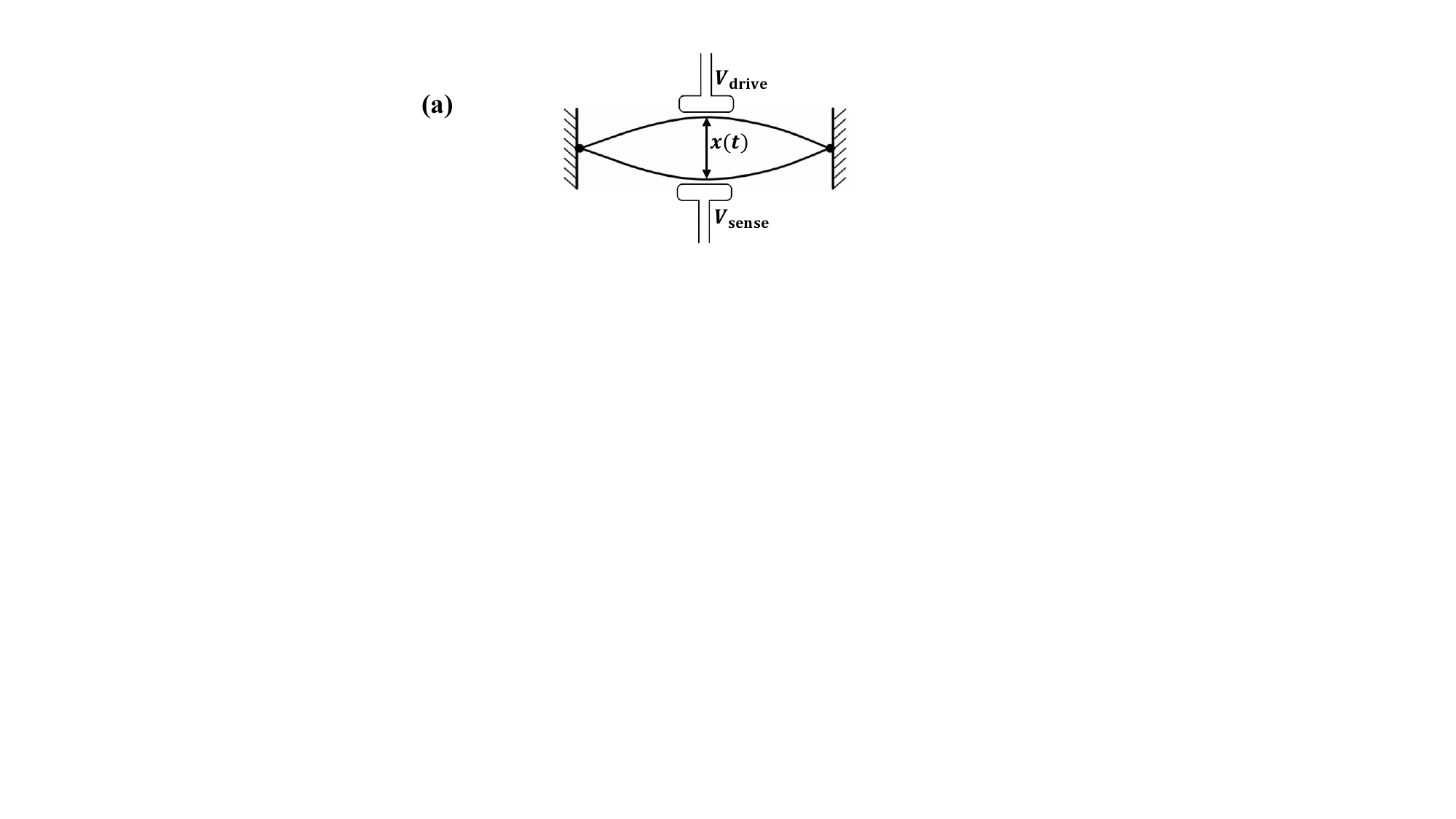}
    % (b) P-Q trajectory
    \includegraphics[
        width=\columnwidth
    ]{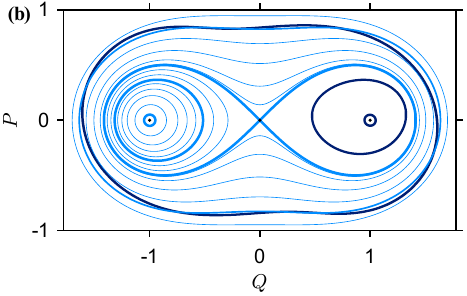}
    \caption{Double-well dynamical system. 
(a) The micromechanical resonator subjected to parametric and direct excitations through electrostatic forcing. The drive electrodes create a parametric drive at frequency $2\omega_1$, and a secondary direct drive at frequency $\omega_2$. The resonator's motion $V_{\mathrm{meas}}(t)$ is measured via an electrical readout signal connected to a lock-in amplifier. The signal is demodulated using homodyne detection, i.e.,  mixing of $V_{\mathrm{sense}}(t)$ with $\sin(\omega_1 t)$ and $\cos(\omega_1 t)$ followed by low-pass filtering, which yields the slowly varying in-phase and quadrature components, $Q(t)$ and $P(t)$. 
(b) Representative double-well dynamics in the $P-Q$ phase portrait, showing theoretical (bright blue) conservative and experimental (dark blue) quasi-conservative dynamics in which the secondary drive is tuned to compensate for energy losses from dissipation.}
    \label{fig:schematic_system}
\end{figure}

Our experimental system is a double-ended tuning-fork (DETF) resonator consisting of two micromechanical beams, each \SI{200}{\micro\meter} long and \SI{6}{\micro\meter} wide. The resonator is electrostatically actuated by two harmonic tones at frequencies $2\omega_1$ (parametric excitation) and $\omega_2$ (direct drive) via a pair of drive electrodes. The resulting motion gives rise to an oscillating voltage at a sensing electrode that is centered between the beams; see Fig. \ref{fig:schematic_system}(a). This voltage is measured with a lock-in amplifier to obtain the in-phase and quadrature components. The resonator dynamics are well described by the following model \cite{margiani2023deterministic}
\begin{align}
\ddot{x}\!+\!2\gamma\dot{x}\!+\!\left[\omega_{0}^{2}\!+\!\lambda\!\cos(2\omega_{1}t)\right]\!x\!+\!\alpha x^{3}=\frac{f}{m}\!\cos(\omega_{2}t),
\label{eq:lab_frame}
\end{align}
where $x(t)\equiv V_{\mathrm{sense}}$ is the modal coordinate measured in volts, $\gamma=\SI{267.8}{\per\second}$ is the linear damping rate, $\omega_{0}/2\pi=\SI{1.1369}{\mega\hertz}$ is the resonator eigenfrequency, and $\alpha=\SI{-4.81e16}{\volt\tothe{-2}\second\tothe{-2}}$ is the Duffing parameter. The parametric drive strength and frequency are given by  $\lambda=\SI{1.782e10}{\second\tothe{-2}}$ and $2\omega_{1}/2\pi=\SI{2.2738}{\mega\hertz}$, respectively. The direct-drive amplitude $f/m$ and frequency $\omega_{2}$ are control parameters that we use to steer the dynamics into or out of chaos (see SM \cite{Shachar2026SM} for details on the calibration of these parameters). 

We restrict our analysis to the dynamics in a frame rotating at $\omega_1$ and spanned by the quadratures $Q$ and $P$ that we define by $x=\ell[Q\cos(\Omega_{1}\tau)-P\sin(\Omega_{1}\tau)]$. Here, $\ell=\sqrt{2\lambda/(3|\alpha|)}$ is a characteristic length scale, $\tau =t\lambda/(4\omega_1)$ is a slow dimensionless time, and $\Omega_1=4\omega_1^2/\lambda$ is the dimensionless parametric drive frequency. When $\gamma/\omega_0$, $\lambda/\omega_0^2$, $\vert\alpha x^2\vert/\omega_0^2$, $f/\vert\omega_0^2x\vert$, $\vert\omega_1-\omega_0\vert/\omega_0$, and $\vert\omega_2-\omega_0\vert/\omega_0$ are all of the order $\ll 1$, Eq. \eqref{eq:lab_frame} can be cast into the following slow-flow equations (see \cite{DYKMAN_&_MARTHALER,Pikovsky_Rosenblum_Kurths_2001,dumont2024energy} and SM \cite{Shachar2026SM})
\begin{align}
\dot Q=\partial_P\mathcal{H}-\varepsilon_Q(\tau),~~ \dot P=-\partial_Q\mathcal{H}-\varepsilon_P(\tau),
\label{eq:q-p_sm}\\
\mathcal{H}=\frac{1-\mu}{2}P^2-\frac{1+\mu}{2}Q^2+\frac{1}{4}(Q^2+P^2)^2. \nonumber
\end{align}

The conservative dynamics of the system is governed by the time-independent dimensionless Hamiltonian $\mathcal{H}(Q,P;\mu)$, which is a conserved quantity of the motion that depends on a single parameter $\mu=4\omega_1(\omega_0-\omega_1)/\lambda$ \cite{DYKMAN_&_MARTHALER}. For $-1<\mu<1$, $\mathcal{H}(Q,P;\mu)$ has the shape of a double well with two equivalent minima at $Q=\pm\sqrt{1+\mu}$, $P=0$, related by the inversion symmetry $(Q,P)\mapsto(-Q,-P)$, and a saddle point at the origin, Fig.~\ref{fig:schematic_system}(b). The curve $\mathcal{H}=0$ that passes through the saddle point $(Q,P)=(0,0)$ is the separatrix, which separates motion confined to one well from motion encircling both wells. The parameter $\mu$ tunes the shape of the double well, and two pitchfork bifurcations at $\mu=\pm1$ create and remove the wells; the cases $\mu<-1$ and $\mu>1$ are described in the SM \cite{Shachar2026SM}. Near either minimum, the conservative motion is periodic with the period $\tau_p=\pi/\sqrt{1+\mu}$, and its exact time dependence follows from elliptic functions (see \cite{DYKMAN_&_MARTHALER} and SM \cite{Shachar2026SM}).

The non-conservative terms $\varepsilon_Q(\tau)=\Gamma Q+F\sin(\Delta_\omega\tau)$ and $\varepsilon_P(\tau)=\Gamma P+F\cos(\Delta_\omega\tau)$ are small perturbations to this conservative dynamics that arise from the presence of damping $\Gamma=4\omega_1\gamma/\lambda$ and the secondary direct drive $F=(f/m)\sqrt{6|\alpha|/\lambda^3}$, where $\Delta_\omega=4\omega_1|(\omega_1-\omega_2)|/\lambda$. These driven dynamics in the rotating frame are very similar to the dynamics of the Duffing-Holmes oscillator. In the presence of non-conservative terms, $\mathcal{H}(Q,P)$ is no longer a constant of the motion and changes as $\mathcal{\dot H}=-(\partial_Q\mathcal{H})\varepsilon_{Q}(\tau) -(\partial_P\mathcal{H})\varepsilon_{P}(\tau)$. 
However, for small perturbations, where $\dot{\mathcal{H}}\propto \varepsilon_{Q,P}\ll1$, we can tune the drive $F$ to compensate for dissipation over one period of oscillation, $\tau_p = 2\pi/\Delta_\omega$. In that case, $\langle\mathcal{\dot H}\rangle_{\tau_p} =(\tau_p)^{-1}\int_0^{\tau_p}\mathcal{\dot H}d\tau=0$ and $Q(\tau)=Q(\tau+\tau_p)$ as well as $P(\tau)=P(\tau+\tau_p)$. In order to achieve this, we find that the drive amplitude needs to satisfy the equation (see SM \cite{Shachar2026SM})
\begin{align}
F\approx\Gamma\frac{\int_0^{\tau_p}[P\dot Q-Q\dot P]d\tau}{\int_0^{\tau_p}[\dot P\sin(\Delta_\omega\tau)-\dot Q\cos(\Delta_\omega\tau)]d\tau}. \label{eq:F_per}
\end{align}
As can be seen from Fig. \ref{fig:schematic_system}(b), these non-conservative periodic trajectories are experimentally accessible and agree remarkably well with the conservative dynamics \footnote{In the experimental measurements, we rotated the $Q$ and $P$ axes with respect to the origin to eliminate damping-induced distortions and clearly show their agreement with the conservative dynamics; see SM \cite{Shachar2026SM} for details.}. The maximal enclosed phase-space area mismatch between the theoretical conservative trajectories and the experimental non-conservative trajectories is less than 5\%; see SM \cite{Shachar2026SM}.

Eq. \eqref{eq:F_per} must be satisfied for any periodic orbits of the non-conservative system. However, \textit{\textbf{not}} all periodic orbits that satisfy Eq. \eqref{eq:F_per} are stable and experimentally accessible. In fact, local orbital stability analysis reveals that the single-well periodic orbits lose their stability via a period-doubling instability that we calculated from Floquet analysis, which we applied to the following pair of linear equations (SM \cite{Shachar2026SM})
\begin{align}
\begin{pmatrix}
\delta \dot Q\\ \delta \dot P
\end{pmatrix}=\begin{pmatrix}-\Gamma+2QP & 1\!-\!\mu\! +\!Q^2\!+\!3P^2\\1\!+\!\mu\!-\!3Q^2\!-\!P^2 & -\Gamma-2QP\end{pmatrix}\begin{pmatrix}
\delta Q\\ \delta  P
\end{pmatrix}\nonumber
\end{align}
and found that one Floquet multiplier crosses -1 at drive values of $F_{\rm pd}=0.248$ and $\Delta_{\omega,{\rm pd}}=2.014$. This is the point where the periodic orbit undergoes a period-doubling bifurcation, giving rise to a new orbit with twice the period. Indeed, as can be shown both by experimental measurements \cite{Eichler_private} and numerical simulations (SM \cite{Shachar2026SM}), the instability of the single-well periodic orbits is part of a period-doubling cascade that culminates in a local single-well chaotic attractor, which can occur in either well due to the inversion symmetry of the conservative dynamics.  
\begin{figure}[t]
    \centering
    \noindent
    \begin{minipage}[c]{0.08\columnwidth}
        \centering
        \rotatebox{90}{$\ell P~(\mathrm{mV})$}
    \end{minipage}%
    \begin{minipage}[c]{0.92\columnwidth}
        \centering
        \noindent
        \begin{minipage}[t]{0.49\linewidth}
            \centering
            \includegraphics[width=\linewidth]{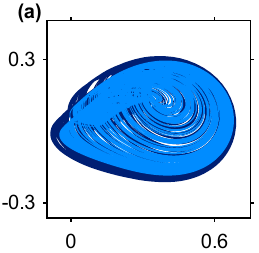}
        \end{minipage}%
        \hfill%
        \begin{minipage}[t]{0.49\linewidth}
            \centering
            \includegraphics[width=\linewidth]{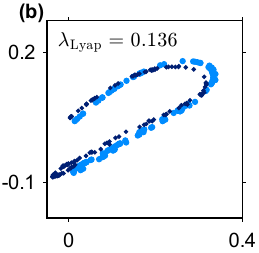}
        \end{minipage}%
        \par\vspace{0.05cm}
        \noindent
        \begin{minipage}[t]{0.49\linewidth}
            \centering
            \includegraphics[width=\linewidth]{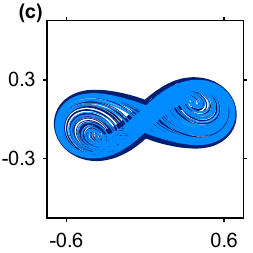}
        \end{minipage}%
        \hfill%
        \begin{minipage}[t]{0.49\linewidth}
            \centering
            \includegraphics[width=\linewidth]{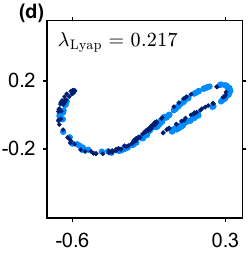}
        \end{minipage}%
        \par\vspace{0.08cm}
        $\ell Q~(\mathrm{mV})$
    \end{minipage}
    \caption{Generated chaotic responses with $\mu=0$ and $\Gamma=0.428$. Panels (a) and (b) show the phase portrait and Poincar\'e map of the single-well chaotic attractor ($F_{\rm pd}<F=0.329<F_{\rm M}=0.330$), and panels (c) and (d) show the phase portrait and Poincar\'e map of the cross-well chaotic attractor ($F=0.339>F_{\rm M}$). The bright blue represents numerical results of Eq. \eqref{eq:q-p_sm} with $\Delta_\omega=2.014$, while dark blue represents experimental measurements. The panels with theoretical and experimental stroboscopic Poincar\'e map $Q(\tau_i)- P(\tau_i)$, where $\tau_i = 2\pi i/\Delta_\omega$, also show the value of the positive (theory-based) Lyapunov exponent of the chaotic attractors.}
    \label{fig:chaotic_motion}
\end{figure}

Applying Eq. \eqref{eq:F_per} to the separatrix trajectory is identical to calculating the zero of the Melnikov function~\cite{guckenheimer2013nonlinear} 
\begin{align}
M(\tau_0)=\int_{-\infty}^{\infty}\mathcal{\dot H}(Q_{\rm sep}(\tau-\tau_0),P_{\rm sep}(\tau-\tau_0))d\tau=0,
\label{eq:Mel_func}
\end{align}
from which we find the critical drive amplitude for global chaos across the two wells, $F_{\rm M}=\Gamma I_1(\mu)/I_2(\mu,\Delta_\omega)$; see SM \cite{Shachar2026SM}. For $\mu=0$, the expression is simplified significantly and reads
\begin{align} 
F_{\rm M}=\frac{2\Gamma}{\pi\Delta_\omega}\ee^{-\pi\Delta_\omega/4}\cosh\left(\frac{\pi\Delta_\omega}{2}\right).
\label{eq:Fcr}
\end{align} 
We note that the expression for $F_{\rm M}$ in Eq. \eqref{eq:Fcr} resembles the result of the classic damped-driven Duffing-Holmes resonator in which $F_{\rm M}=[4\Gamma/(3\sqrt{2}\pi\Delta_\omega)]\cosh(\pi\Delta_\omega/2)$. Fig.~~\ref{fig:chaotic_motion} illustrates the transition from localized to cross-well chaos. Below the Melnikov threshold, the phase portrait remains confined to a single well [Fig.~~\ref{fig:chaotic_motion}(a)], while the stroboscopic Poincaré map resolves narrow, curved bands [Fig.~~\ref{fig:chaotic_motion}(b)]. The positive numerical Lyapunov exponent, $\lambda_{\rm Lyap}=0.136$, confirms the chaotic nature of this localized motion. Above the threshold, trajectories explore both wells [Fig.~~\ref{fig:chaotic_motion}(c)], and the Poincaré map develops folded bands extending across both signs of $Q$ [Fig.~~\ref{fig:chaotic_motion}(d)], with $\lambda_{\rm Lyap}=0.217$. This global transport is enabled by transverse intersections of the stable and unstable manifolds of the saddle periodic orbit, predicted by the Melnikov analysis for $F>F_{\rm M}$. The resulting homoclinic tangle stretches nearby initial conditions along the unstable direction and contracts them along the stable direction, while excursions around the wells fold material segments back toward the saddle; see SM \cite{Shachar2026SM}. Repeated stretching, folding, and return generate fine structures in phase space, allowing chaotic trajectories to visit both wells. Numerical and experimental results exhibit closely matching phase portraits and Poincaré maps, with the observed transition bracketed by $F=0.329$ and $F=0.339$, consistent with the prediction $F_{\rm M}=0.330$.

\begin{figure}[t]
    \centering
    \begin{minipage}{\columnwidth}
        \centering
        \includegraphics[width=\linewidth]{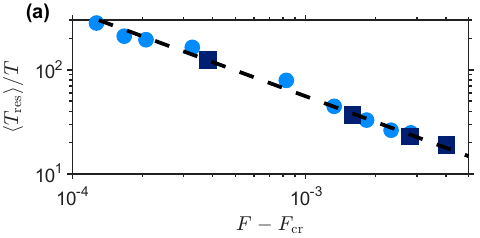}
    \end{minipage}
    \vspace{0.08cm}
\noindent
\begin{minipage}[c]{0.035\columnwidth}
    \centering
    \raisebox{0.30cm}{%
        \rotatebox{90}{$\ell Q~(\mathrm{mV})$}%
    }
\end{minipage}%
\begin{minipage}[c]{0.965\columnwidth}
    \centering
    % (b) 70.25 mV
    \includegraphics[
        width=\linewidth,
        trim=0cm 0.45cm 0cm 0cm,
        clip
    ]{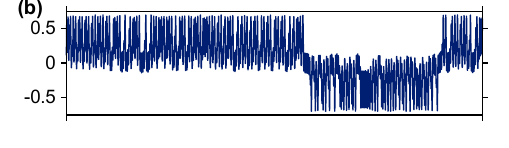}
   
    % (c) 71 mV
    \includegraphics[
        width=\linewidth,
        trim=0cm 0.05cm 0cm 0.00cm,
        clip
    ]{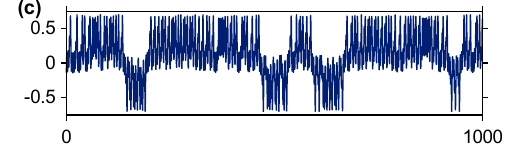}

    $\tau$
\end{minipage}
    \caption{Residence-time scaling near the onset of global chaos. (a) Mean residence time between successive interwell transitions, normalized by the forcing period $T$, as a function of the distance from the critical forcing amplitude, $F-F_{\mathrm{cr}}$. Bright blue circles and dark blue squares represent numerical and experimental results, respectively, plotted relative to their respective estimated critical forcing amplitudes, $F_{\mathrm{cr}}^{\mathrm{num}}=0.33367$ and $F_{\mathrm{cr}}^{\mathrm{exp}}=0.33987$ ($70.17\,\mathrm{mV}$). The dashed line is the theoretical power-law $\langle T_{\mathrm{res}}\rangle/T\propto(F-F_{\mathrm{cr}})^{-\eta}$, with the exponent $\eta=0.824$ calculated from the Floquet multipliers of the saddle periodic orbit evaluated at the Melnikov-predicted threshold. The coefficient of determination of the numerical and experimental data with the theoretical power law is $R^2=0.918$. (b) Experimental time series of the rotating-frame quadrature at a driving amplitude of $70.25\,\mathrm{mV}$, corresponding to the leftmost dark blue square in (a). (c) Experimental time series at a driving amplitude of $71\,\mathrm{mV}$, corresponding to the rightmost dark blue square in (a).}
    \label{fig:scaling_law}
\end{figure}

We identify that the transition from a localized single-well chaotic attractor to a global double-well chaotic attractor proceeds via a merging crisis when we analyze how the average time of chaotic switching between the wells varies as a function of $F-F_{\rm cr}$ (Fig.~\ref{fig:scaling_law}). In particular, near the onset of cross-well global chaos, trajectories exhibit long episodes of chaotic motion confined to either well, interrupted by transitions between the two. This behavior is characteristic of crisis-induced intermittency, for which the mean residence time between successive transitions is expected to follow a power law with the distance from the critical forcing amplitude, $\langle T_{\mathrm{res}}\rangle\propto(F-F_{\mathrm{cr}})^{-\eta}$ \cite{grebogi1987critical}. The experimental measurements exhibit a pronounced decrease in residence time as forcing increases, consistent with this behavior. Moreover,  a theoretical Floquet-based calculation of the exponent yields $\eta=0.824$ (see SM \cite{Shachar2026SM}), which agrees remarkably well with both experimental measurements and numerical simulations [Fig.~\ref{fig:scaling_law}(a)], where $\eta_{\rm exp}/\eta=0.990,~\eta_{\rm num}/\eta=1.082$.

Our results demonstrate how the character of chaos evolves in a driven nonlinear system. Our demonstration is based on a parametrically driven nonlinear resonator, whose rotating-frame dynamics resemble those of a particle in a double-well potential. A saddle point separates local inner-well from global double-well oscillations. Below the Melnikov threshold, the saddle point acts as a barrier, and the chaotic motion remains confined to one well. The barrier breaks down when the stable and unstable manifolds of the saddle point intersect in a merging crisis. The agreement between a Floquet-based prediction of the scaling exponent and the statistics of global chaotic motion links the local instability of a mediating periodic orbit to those statistics. Interestingly, we find that a closed-form criterion from a conservative, near-resonant theory predicts the merging threshold in a resonator with a quality factor of only about $10^4$~\cite{agarwal2006optimal,polunin2016characterization,zega2020numerical}, showing that the criterion is robust against dissipation. Because many weakly nonlinear oscillators reduce to the same rotating-frame description, the picture is likely to apply to a range of physical systems beyond micromechanical resonators. Together, these findings establish an experimentally accessible route to investigating and tailoring the extent and switching timescales of chaos in driven nonlinear resonators. This is highly beneficial to various chaos-based applications ~\cite{tsonis2012chaos,akgul2016chaos}, including random number generation~\cite{akashi2022mechanical,quinn2023random,wei2026true}, weak signal detection~\cite{hu2003application,yan2023application,cheng2023analysis}, and the design and generation of signals for secure communication and signal processing~\cite{Chaotic_signals_and_signal_processing,kaddoum2016wireless,hwang2023machine,zhang2023chaos}.

\begin{acknowledgments}

\end{acknowledgments}
%\appendix
\begin{center}{\bf Data Availability Statement}
\end{center}
The data that support the findings of this study are available from the corresponding author upon reasonable request.

\bibliography{Amitbib}

\end{document}

\title{\textbf{Supplemental Material}\\
From Local to Global Chaos in a Parametric Micromechanical Resonator}

\author{Amit Shachar, Gabriel Margiani,
Alexander Eichler, Oriel Shoshani}

\date{}

\maketitle

\vspace{0.5cm}

\setcounter{tocdepth}{2}
\tableofcontents

\newpage

\section{ System identification and parameter estimation}
\label{sec:experiment}
\subsection{Estimation of eigenfrequency and linear damping}
\begin{figure}[htbp]
    \centering
    \includegraphics[width=0.6\textwidth]
    {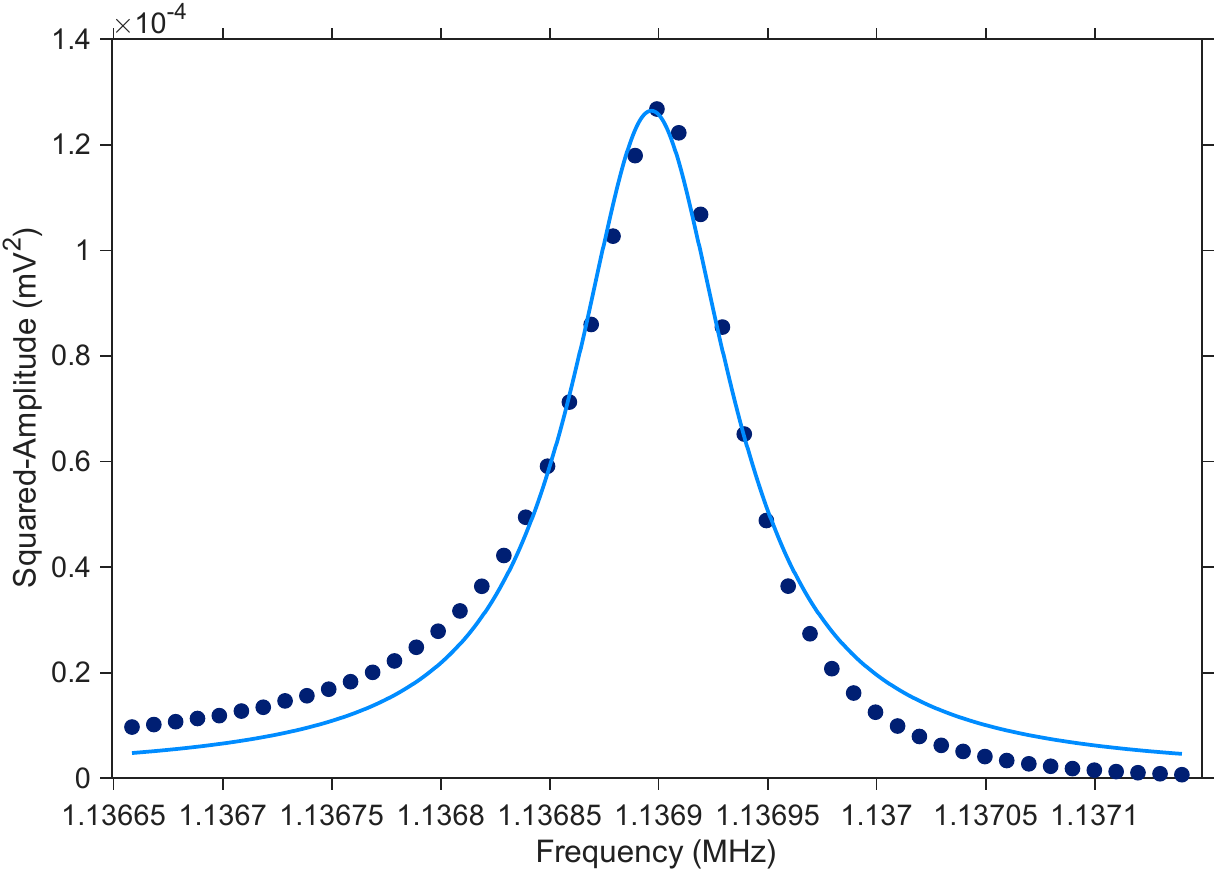}
    \caption{Linear response of the DETF resonator. The squared response amplitude is shown as a function of the direct-drive frequency. Dark blue points represent experimental measurements, while the bright blue curve represents the fitted Lorentzian response. The fit yields $f_0=1.136906\,\mathrm{MHz}$ and $\gamma\simeq267.8\,\mathrm{s}^{-1}$, corresponding to $Q=13334\pm28$.}
    \label{fig:direct-response-fit}
\end{figure}

The eigenfrequency and linear damping were determined by measuring the resonator response under weak direct excitation. The measured squared amplitude was fitted to a Lorentzian function, from which we determined the eigenfrequency and the half-width at half-maximum (HWHM) that equals the linear damping parameter $\gamma$. Fig.~\ref{fig:direct-response-fit} shows the measured squared amplitude and the fitted Lorentzian. The fit yields an eigenfrequency of $f_0=1.1369\,\mathrm{MHz}$, and HWHM of $\gamma=267.8\,\mathrm{s}^{-1}$, which in turn reslts in a quality factor of $Q=13334\pm28$. 
frequency.

\FloatBarrier
\subsection{Parametric excitation and instability threshold}
\begin{figure}[htbp]
    \centering
    \includegraphics[width=0.6\textwidth]
    {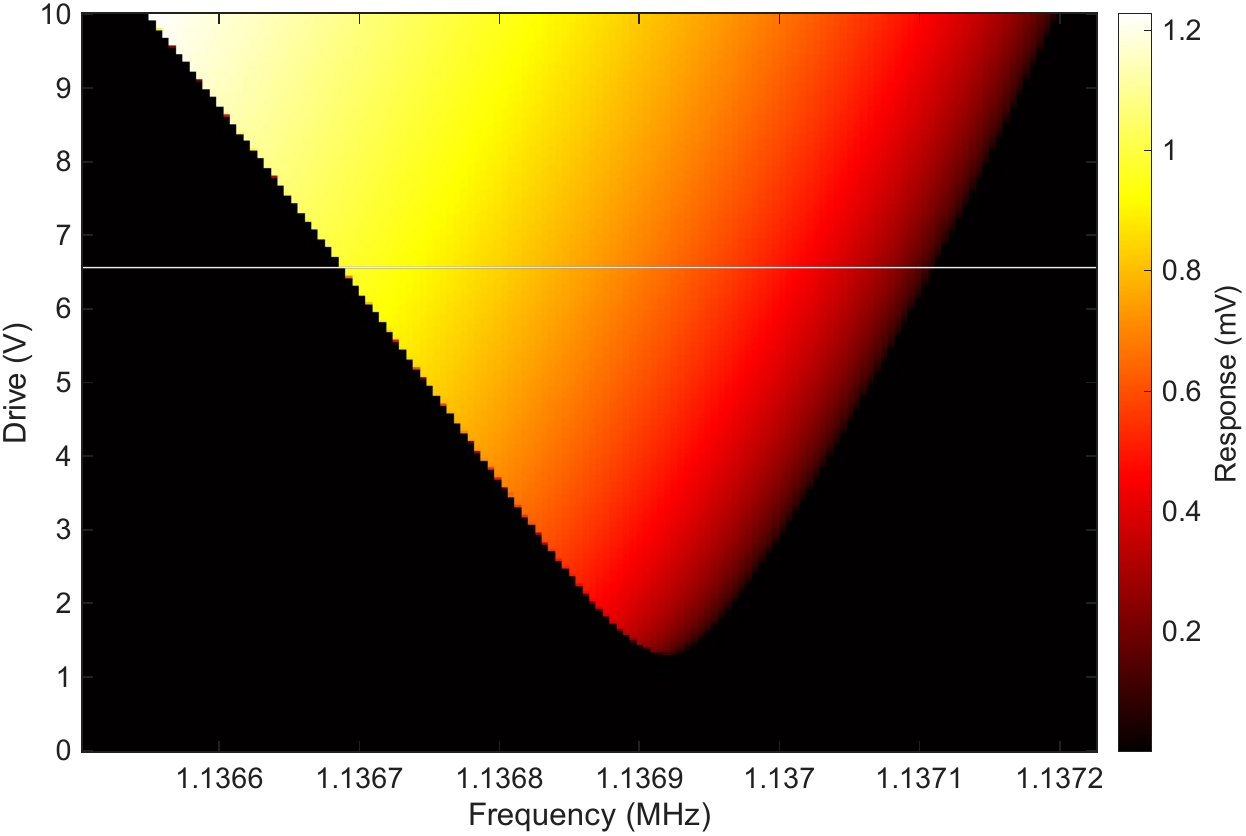}
    \caption{Experimental Arnold tongue of the parametrically excited DETF resonator. The color scale represents the measured oscillation amplitude as a function of the parametric drive amplitude and half the parametric excitation frequency. The instability region emerges above the threshold voltage $U_{\mathrm{th}}=(1.288\pm0.0089)\,\mathrm{V}$ and broadens with increasing parametric excitation.}
    \label{fig:arnold_detf}
\end{figure}

The parametric excitation strength was characterized by measuring the resonator response as a function of the parametric driving frequency and amplitude, in the absence of an additional direct drive. Fig.~\ref{fig:arnold_detf} presents the resulting parametric instability region, commonly known as the Arnold tongue. Below the instability threshold, the resonator remains in its trivial state. Above the threshold, parametric excitation overcomes linear dissipation, producing a finite oscillation amplitude. The instability region widens as the parametric drive amplitude increases. The measured threshold voltage at the first parametric instability is
\begin{align}
U_{\mathrm{th}}=(1.288\pm0.0089)\,\mathrm{V}.
\end{align}
For excitation near twice the natural frequency, the threshold condition is
\begin{align}
\lambda=\frac{2U_{\mathrm{param}}}{Q\cdot U_{\mathrm{th}}}\omega_0^2=1.782\times10^{10} \,\mathrm{s^{-2}} ,
\end{align}
where $U_{\mathrm{param}}=3\,\mathrm{V}$ is the parametric voltage strength that is used in the experiment and $\lambda$ is the parametric modulation coefficient appearing in the equation of motion. This relation provides the calibration between the applied parametric drive voltage and the modulation strength used in the theoretical model.

\FloatBarrier
\subsection{Estimation of Duffing parameter and nonlinear damping}

\begin{figure}[htbp]
    \centering
    \includegraphics[width=0.6\textwidth]{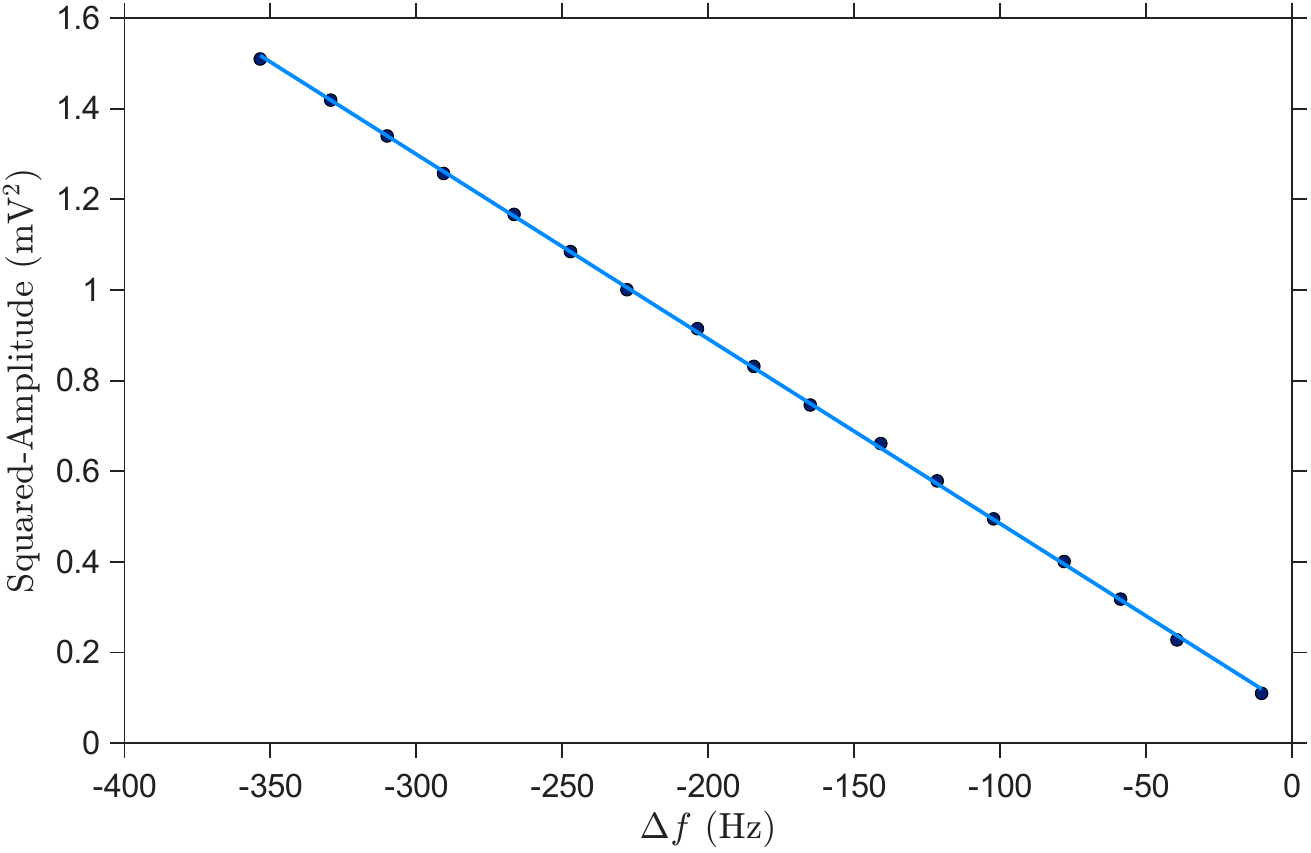}
    \caption{
    Experimental Duffing backbone of the DETF resonator. The squared peak amplitude, $a_{\rm peak}^2$, is plotted as a function of the resonance-frequency shift, $\Delta f=f_{\mathrm{peak}}-f_0$, extracted from direct-drive frequency sweeps at different driving amplitudes. Black circles represent the experimental data, and the blue solid line represents the linear fit. The negative frequency shift with increasing amplitude indicates a softening Duffing nonlinearity, characterized by $\alpha=-4.81\times10^{16}\,\mathrm{V}^{-2}\mathrm{s}^{-2}$.}
    \label{fig:duffing_detf_2}
\end{figure}

\begin{figure}[htbp]
    \centering
    \includegraphics[width=0.6\textwidth]{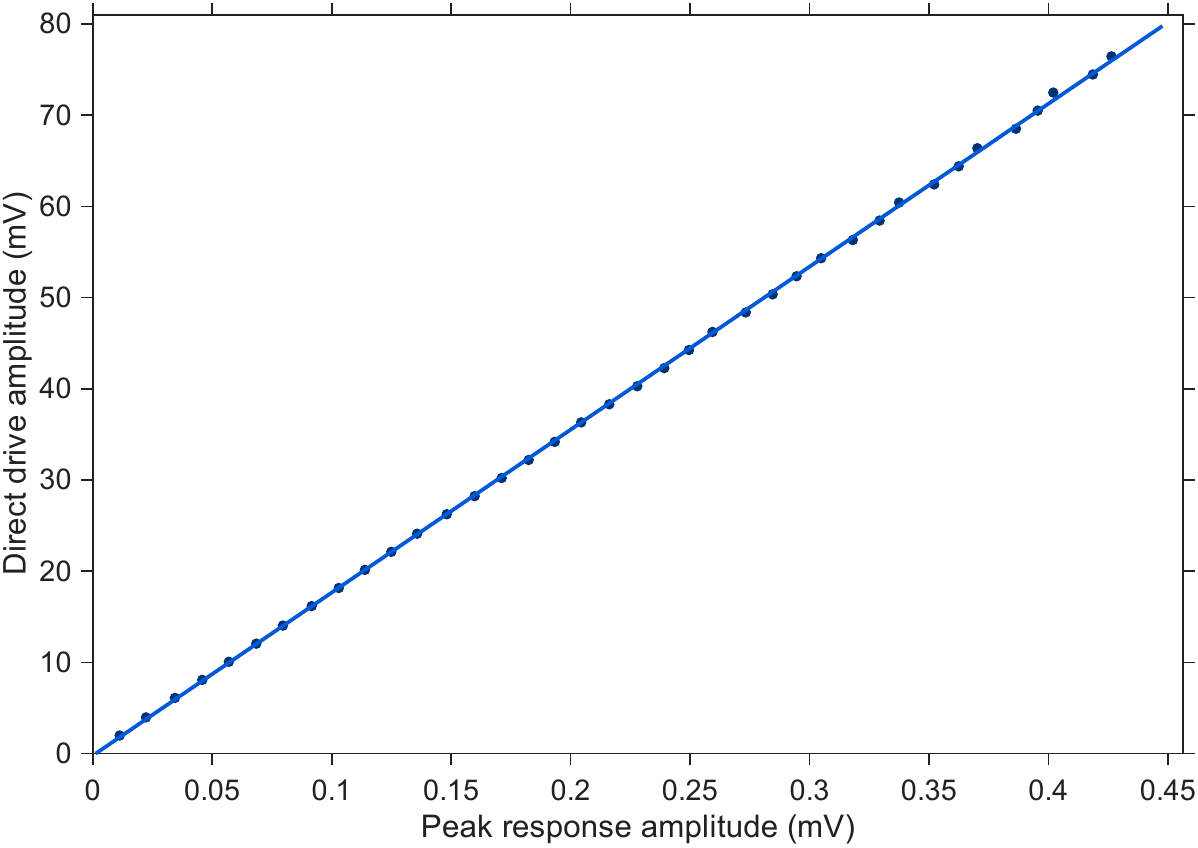}
    \caption{
Direct-drive amplitude response of the DETF resonator for applied driving voltages up to $78\,\mathrm{mV}$. The dark-blue circles represent the experimentally measured peak amplitudes, while the solid blue line represents the linear fit. The approximately linear dependence of the peak response on the driving amplitude, together with the negligible improvement obtained by including nonlinear damping in the full frequency-response fit, indicates that nonlinear damping is not required to describe the direct-drive response within the investigated range.
}
\label{fig:nonlinear_damping}
\end{figure}

The conservative cubic (Duffing/Kerr) nonlinearity of the resonator was characterized by measuring the frequency-squared amplitude backbone curve. The backbone curve of the Duffing resonator is given by $\Omega-\omega_0=3\alpha a_{\rm peak}^2/(8\omega_0)$, and denotes the location of the peak amplitude of the response curve $(a_{\rm peak})$ \cite{lifshitz2008nonlinear}.  To quantify this effect, we measured the driven response curve for various direct-drive strengths and extracted the frequency and the corresponding peak amplitude for each frequency sweep. The squared peak amplitude was then plotted as a function of the frequency shift, $\Delta f=f_{\mathrm{peak}}-f_0$, where $f_0$ is the natural resonance frequency. As shown in Fig.~\ref{fig:duffing_detf_2}, the squared amplitude exhibits an approximately linear dependence on the frequency shift, consistent with the Duffing backbone relation. The negative slope indicates a negative Duffing coefficient (softening nonlinearity). Fitting the experimental data yields $\alpha=-4.81\times10^{16}\,
\mathrm{V}^{-2}\mathrm{s}^{-2}$.
The negative sign of $\alpha$ is consistent with the experimentally observed softening behavior and is incorporated into the rotating-frame model. In the next section (Sec.~\ref{sec:rotating_frame}), we will use its absolute value, $|\alpha|$.
\FloatBarrier
For the nonlinear damping estimation, we measured the direct-drive amplitude as a function of the resonator peak amplitude. We note that for the Duffing resonator this relation is given by $f/(2\omega_0m)=a_{\rm peak}(\gamma+\gamma_{\rm nl}a_{\rm peak}^2/4)$, where the nonlinear damping corresponds to a van der Pol damping \cite{lifshitz2008nonlinear}, i.e., an additional damping term of the form $2\gamma_{\rm nl}x^2\dot x$ in Eq. (1) of the main text. As shown in Fig.~\ref{fig:nonlinear_damping}, the measured relation between the direct-drive amplitude and the resonator peak amplitude is approximately linear within the investigated range. Therefore, nonlinear damping can be neglected in the simplified theoretical description of the direct-drive response within this range.
\FloatBarrier
\subsection{Direct-drive amplitude calibration}
\begin{figure}[htbp]
    \centering
    \includegraphics[width=0.6\textwidth]
    {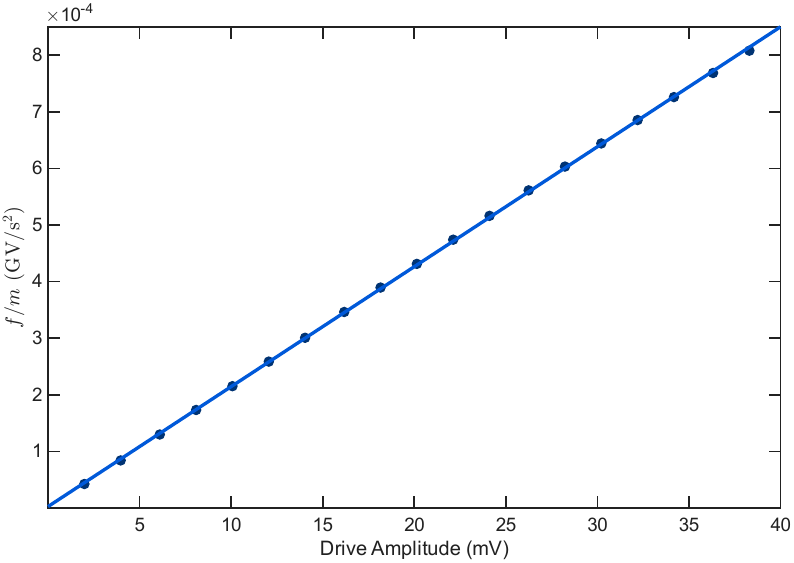}
    \caption{Calibration of the effective direct-drive forcing amplitude $f/m$ as a function of the applied driving voltage $U_d$. Blue points represent the forcing amplitudes extracted from the experimental linear-response measurements, while the orange line denotes a linear fit. The fitted slope is $k_d=2.14581\times10^7\,\mathrm{s}^{-2}$, establishing the conversion between the experimental voltage and the forcing amplitude used in the theoretical model.}
    \label{fig:drive_to_f0m}
\end{figure}

To compare the experimental measurements with the theoretical rotating-frame model, the applied direct-drive voltage must be converted into the corresponding effective forcing amplitude $f/m$. This conversion was obtained by fitting the measured linear oscillator response at different applied drive amplitudes. The fitted forcing amplitude exhibits an approximately linear dependence on the applied voltage, as shown in Fig.~\ref{fig:drive_to_f0m}.
The resulting calibration can be expressed as
\begin{align}
\frac{f}{m}=k_d U_d+b_d,
\end{align}
where $U_{\mathrm{drive}}$ is the applied direct-drive voltage, $k_d$ is the voltage-to-force calibration coefficient, and $b_d$ is the fitted offset. The linear fit yields
\begin{align}
k_d=(2.14581\pm0.00069)\times10^7\,
\mathrm{s}^{-2},
\end{align}
when $U_{\mathrm{drive}}$ is expressed in volts and $f/m$ in $\mathrm{V\,s}^{-2}$. The fitted offset is $b_d=(-482\pm157)\,\mathrm{V\,s}^{-2}$. This calibration provides the correspondence between the experimentally controlled drive voltage and the dimensional forcing amplitude, which is subsequently converted into the dimensionless parameter $F$ of the rotating-frame equations.

\FloatBarrier
\section{ Derivation of the rotating-frame equations}
\label{sec:rotating_frame}

The dynamics of the nonlinear parametrically driven resonator
with an additional weak direct drive are described by Eq. (1) of the main text.
To derive the slow evolution equations of the in-phase and quadrature components in the rotating frame, we substitute the ansatz $x=A(t)\ee^{\ii\omega_1 t}+A^*(t)\ee^{-\ii\omega_1 t}$ into Eq. (1) of the main text,
where $A(t)$, the complex amplitude, and its complex conjugate $A^*(t)$, satisfy the constraint $\dot A\ee^{\ii\omega_1 t}+\dot A^*\ee^{-\ii\omega_1 t}=0$, such that $\dot x= \ii\omega_1 A\ee^{\ii\omega_1 t}-\ii\omega_1 A^*\ee^{-\ii\omega_1 t}$. Computing all the terms in Eq. (1) using the above ansatz and constraint on the complex amplitude, and removing the fast-oscillating terms at $\ee^{\pm 3\ii\omega_1t}$, we get the following complex-amplitude equation

\begin{align*}
\dot A
=&-\ii\delta_\omega A-\gamma A
+\frac{3\ii\alpha}{2\omega_1}|A|^2A
+\frac{\ii\lambda}{4\omega_1}A^*
-\frac{\ii f}{4m\omega_1}\ee^{\ii\Theta}.
\end{align*}

where $\delta_\omega=\omega_1-\omega_0$, and $\Theta=(\omega_2-\omega_1)t$. Using Cartesian notation $A=(q+\ii p)/2$, we obtain the following pair of  evolution equations for the in-phase and quadrature components

\begin{align}
\dot q &=
\delta_\omega p-\gamma q
-\frac{3\alpha}{8\omega_1}(q^2+p^2)p
+\frac{\lambda}{4\omega_1}p
+\frac{f}{2m\omega_1}\sin\Theta,
\\
\dot p &=
-\delta_\omega q-\gamma p
+\frac{3\alpha}{8\omega_1}(q^2+p^2)q
+\frac{\lambda}{4\omega_1}q
-\frac{f}{2m\omega_1}\cos\Theta.
\end{align}

\subsection{Dimensional Hamiltonian and scaling}
The time-independent dimensional Hamiltonian is given by
\begin{align}
H=\frac{\delta_\omega}{2} (q^2+p^2)-
 \frac{3\alpha}{32\omega_1}(q^2+p^2)^2-\frac{\lambda}{8\omega_1}(q^2-p^2),
\end{align}
and its non-dimensional counterpart $\mathcal{H}$ can be derived as follows. We define the following non-dimensional quadrature $q=\ell Q$ and $p=\ell P$, where $\ell$ is the yet to be determined characteristic length scale. Hence, the dimensional Hamiltonian reads
\begin{align}
H=\frac{\delta_\omega}{2} \ell^2(Q^2+P^2)-
\frac{3\alpha}{32\omega_1}\ell^4(Q^2+P^2)^2-\frac{\lambda}{8\omega_1}\ell^2(Q^2-P^2).
\label{eq:H_scale}
\end{align}
Therefore, dividing Eq. \eqref{eq:H_scale} by $\ell^2$ and multiplying by the yet to be determined characteristic time scale $\mathcal{T}$, we find that
\begin{align*}
\mathcal{H}\equiv\mathcal{T}\frac{H}{\ell^2}=\frac{1}{2}Q^2 [\mathcal{T}(\delta_\omega-\frac{\lambda}{4\omega_1})]-
\frac{3\alpha}{32\omega_1}\mathcal{T}\ell^2(Q^2+P^2)^2+\frac{1}{2}P^2[\mathcal{T}(\delta_\omega+\frac{\lambda}{4\omega_1})].
\end{align*}
The choice of $\ell$ and $\mathcal{T}$ is not unique; however, following Ref. \cite{DYKMAN_&_MARTHALER}, we use
the constraint equations 
\begin{align*}
\mathcal{T}\ell^2\frac{-3\alpha}{32\omega_1}=\frac{1}{4},~
\mathcal{T}\frac{\lambda}{4\omega_1}=1, 
\end{align*}
obtaining the characteristic time $\mathcal{T}$ and length $\ell$ 
\begin{align}
\mathcal{T} =4\omega_1/\lambda,~
\ell=\sqrt{2\lambda/(-3\alpha)}.
\end{align}
Hence, the dimensionless Hamiltonian $\mathcal{H}$ contain a single dimensionless parameter $\mu=4\omega_1(\omega_0-\omega_1)/\lambda$ and given by 
\begin{align}
\mathcal{H}=-\frac{1}{2}Q^2 (1+\mu)+\frac{1}{2}P^2(1-\mu)+
 \frac{1}{4}(Q^2+P^2)^2.
 \label{eq:q-p_sm}
\end{align}

\subsection{Dimensionless equations with non-conservative perturbations}
In the rotating frame, the slow dynamics of $Q$ and $P$ are given by
\begin{align}
\dot Q
&=
\partial_P \mathcal{H}
-
\Gamma Q
-
F\sin(\Delta_\omega\tau),
\label{eq:Q_rotating_frame}\\
\dot P
&=
-\partial_Q \mathcal{H}
-
\Gamma P
-
F\cos(\Delta_\omega\tau),
\label{eq:P_rotating_frame}
\end{align}
where
\begin{align}
\Gamma=\frac{4\omega_1\gamma_l}{\lambda},~
F=
\frac{f}{m}
\sqrt{\frac{-6\alpha}{\lambda^3}},~
\Delta_\omega=
\frac{4\omega_1(\omega_1-\omega_2)}{\lambda}.
\end{align}

% ============================================================

\section{ Closed-form conservative dynamics}
\label{sec:analytic_trajectories}

For $\Gamma=F=0$, the Hamiltonian is a conserved quantity, i.e., $d\mathcal{H}/d\tau=0$. Therefore, the $Q-P$ dynamics can be solved in closed form. Introducing the canonical transformation 
\begin{align}
I=\frac{Q^2+P^2}{2},
\qquad
\theta=\tan^{-1}\left(\frac{Q}{P}\right),
\end{align}
the Hamiltonian becomes $\mathcal{H}=I\left(I-\mu+\cos 2\theta\right)$.
The conservative dynamics of $I$ and $\theta$ are governed by the following pair of evolution equations  

\begin{align}
\dot I &=-\frac{\partial\mathcal H}{\partial\theta}
=2I\sin(2\theta),
\label{eq:Ism}\\
\dot\theta &=\frac{\partial\mathcal H}{\partial I}
=2I-\mu+\cos(2\theta).
\label{eq:theta}
\end{align}

By squaring and adding Eq.\eqref{eq:Ism} and $\mathcal{H}$, we effectively decouple the $I$ dynamics from its $\theta$ dependency, and obtain the following single equation for the evolution of $I$
\begin{equation}
\frac{\dot I^2}{4}+U_{\rm eff}(I)=0,
\label{eq:IevPot}
\end{equation}
where $U_{\rm eff}(I)$ can be viewed as an effective potential, which is given by
\begin{align}
U_{\rm eff}(I)=\left(\mathcal{H}_i-I^2+\mu I\right)^2-I^2=[\mathcal{H}_i-I^2+(\mu-1)I][\mathcal{H}_i-I^2+(\mu+1)I].
\label{eq:U_eff}
\end{align}
The dynamics of Eq. \eqref{eq:IevPot} can be interpreted as the motion of a particle with a coordinate $I$ trapped in a potential well $U_{\rm eff}(I)$. The shape of the potential well $U_{\rm eff}(I)$ depends on the parameter $\mu$, and the value of the Hamiltonian $\mathcal{H}_i$, which is determined by the initial conditions of  $I$ and $\theta$. In particular, the value of the Hamiltonian $\mathcal{H}_i$ determines the turning points $I_1\geq I_{2}$ at which $\dot I=0$. These turning points are the two positive and real roots of the equation $U_{\rm eff}(I)=0$, which depend on the values of $\mu$ and $\mathcal{H}_i$. Specifically, for $\mu<-1$, the Hamiltonian levels are positive definite $\mathcal{H}_i\geq0$, and the turning points are given by $I_{1}=(\mu+1)/2 + \sqrt{[(\mu+1)/{2}]^2+\mathcal{H}_i}$ and $I_{2}=(\mu-1)/2 + \sqrt{[(\mu-1)/{2}]^2+\mathcal{H}_i}$, where the other two roots $I_{3}=(\mu+1)/2 - \sqrt{[(\mu+1)/{2}]^2+\mathcal{H}_i}$ and $I_{4}=(\mu-1)/2 - \sqrt{[(\mu-1)/{2}]^2+\mathcal{H}_i}$  are negative, and therefore, non-physical, since $I=(Q^2+P^2)/2\geq0$. For $\mu\geq-1$, the Hamiltonian levels are bounded from below by the value $\mathcal{H}_{\rm min}=-(\mu+1)^2/4\leq\mathcal{H}_i$, and hence, for $-(\mu+1)^2/4\leq\mathcal{H}_i<-(\mu-1)^2/4$,  the turning points are $I_{1,2}=(\mu+1)/2 \pm \sqrt{[(\mu+1)/{2}]^2+\mathcal{H}_i}$, where the two other roots $I_{3,4}=(\mu-1)/2\pm\sqrt{[(\mu-1)/{2}]^2+\mathcal{H}_i}$ are complex-conjugates $I_3=I_4^*$, and for $\mathcal{H}_i\geq-(\mu-1)^2/4$,  the turning points are $I_{1}=(\mu+1)/2 + \sqrt{[(\mu+1)/{2}]^2+\mathcal{H}_i}$ and $I_{2}=(\mu-1)/2 + \sqrt{[(\mu-1)/{2}]^2+\mathcal{H}_i}$, where the other two roots $I_{3}=(\mu+1)/2 - \sqrt{[(\mu+1)/{2}]^2+\mathcal{H}_i}$ and $I_{4}=(\mu-1)/2 - \sqrt{[(\mu-1)/{2}]^2+\mathcal{H}_i}$  are negative.
For $I_3=I_4^*$, the time dependence of $I=I(\tau)$ in terms of the Jacobi elliptic functions is given by (see Ref. \cite{byrd2013handbook}, pp 133)
\begin{align}
I(\tau)=\frac{z_2I_1+z_1I_2-(z_2I_1-z_1I_2){\rm cn}[\Omega(\tau-\tau_0)|m]}{z_1+z_2-(z_2-z_1){\rm cn}[\Omega(\tau-\tau_0)|m]},
\label{eq:I(t)op1}
\end{align}
where $\Omega=2\sqrt{z_1z_2}$, $z_j=\sqrt{(I_j-I_3)(I_j-I_4)}$ with $(j=1,2)$, and the parameter of the elliptic function $m_{\mathcal{H}_i}=[(I_1-I_2)^2-(z_1-z_2)^2]/(4z_1z_2)$ and $\tau_0$ are set by the initial conditions (along with the value of the Hamiltonian $\mathcal{H}_i$). For $I_4<I_3<0$, the time dependence of $I=I(\tau)$ in terms of the Jacobi elliptic functions is given by (see Ref. \cite{byrd2013handbook}, pp 128)
\begin{align}
I(\tau)=
\frac{
I_1(I_2-I_4)+I_4(I_1-I_2)
{\rm sn}^2\left[\Omega(\tau-\tau_0)|m\right]
}{
(I_2-I_4)+(I_1-I_2)
{\rm sn}^2\left[\Omega(\tau-\tau_0)|m\right]
},
\label{eq:I(t)op2}
\end{align}
where $\Omega=\sqrt{(I_1-I_3)(I_2-I_4)}$, and the parameter of the elliptic function $m_{\mathcal{H}_i}=(I_1-I_2)(I_3-I_4)/[(I_1-I_3)(I_2-I_4)]$ and $\tau_0$ are set by the initial conditions (along with the value of the Hamiltonian $\mathcal{H}_i$). 

For a known $I(\tau)$, we can calculate $Q(\tau)$ and $P(\tau)$ from Eq. \eqref{eq:Ism} and $\mathcal{H}(I,\theta)$, which can be rewritten as 
\begin{align}
\dot I=2QP,~\mathcal{H}_i-I^2+\mu I=\frac{P^2-Q^2}{2}.
\label{eq:q&p_con}
\end{align}
These two equations have up to four possible solutions for $Q$ and $P$. Nevertheless, from the second equation in Eq. \eqref{eq:q&p_con} and the definition of $I$, we arrive at the following pair of equations 
$$ Q^2(\tau)=I^2(\tau)+(1-\mu)I(\tau)-\mathcal H_i,~P^2(\tau)=-I^2(\tau)+(1+\mu)I(\tau)+\mathcal H_i,$$
which can then be used along with the first equation in Eq. \eqref{eq:q&p_con}, to find the appropriate $P(\tau)$ for each of the pairs of possible solutions of $Q(\tau)$.
Moreover, in our case of interest, where $-1<\mu<1$, the expressions can be simplified such that for inner-well trajectories $(\mathcal{H}_i<0)$,
\begin{align}
&Q(\tau)=\pm\frac{2^{3/2}|\mathcal{H}_i|^{1/2}\operatorname{dn}(\tau'-\tau'_0)}{\kappa_{+}+\kappa_{-}\operatorname{cn}(\tau'-\tau'_0)},\quad
&P(\tau)=\frac{\kappa_{+}\kappa_{-}|\mathcal{H}_i|^{1/4}\operatorname{sn}(\tau'-\tau'_0)}{\kappa_{+}+\kappa_{-}\operatorname{cn}(\tau'-\tau'_0)},
\label{eq:Q-P_vs_t}
\end{align}
where $\kappa_{\pm}=(1+\mu\pm2|\mathcal{H}_i|^{1/2})^{1/2}$, $\tau'=2^{3/2}|\mathcal{H}_i|^{1/4}\tau$, and $-(1+\mu)^2/4\leq\mathcal{H}_i<0$. The period of oscillation is given by $\tau_p=\sqrt{2/\sqrt{|\mathcal{H}_i|}}K(m_{\mathcal{H}_i})$, where $K(m_{\mathcal{H}_i})$ is the complete elliptic integral of the first kind with the parameter $m_{\mathcal{H}_i}=[\mu^2-(1-2|\mathcal{H}_i|^{1/2})^2]/(8|\mathcal{H}_i|^{1/2})$. For negative values of the parameter $m_{\mathcal{H}_i}<0$, the functions ${\rm cn}(\tau'|m_{\mathcal{H}_i}), ~{\rm sn}(\tau'|m_{\mathcal{H}_i})$, and ${\rm dn}(\tau'|m_{\mathcal{H}_i})$, need to be replaced with ${\rm cd}(\tilde\tau'|\tilde m_{\mathcal{H}_i}),~{\rm sd}(\tilde\tau'|\tilde m_{\mathcal{H}_i})/\sqrt{1-m_{\mathcal{H}_i}}$, and ${\rm nd}(\tilde\tau'|\tilde m_{\mathcal{H}_i})$, respectively, where $\tilde m_{\mathcal{H}_i}=-m_{\mathcal{H}_i}/(1- m_{\mathcal{H}_i})$, and $\tilde\tau'=\tau'\sqrt{1-m_{\mathcal{H}_i}}$.

\subsection{The separatrix trajectory}
\label{subsec:separatrix}
On the separatrix $\mathcal{H}_i=0$; hence, the non-trivial $(I\neq0)$ separatrix trajectory satisfy the equation  
\begin{align}
I=\mu-\cos2\theta.
\label{eq:I_and_theta_new}
\end{align}
Therefore, we find from Eq. \eqref{eq:theta} that
\begin{align}
\dot{\theta}=\mu-\cos2\theta.
\label{eq:theta_dot_action}
\end{align}
To integrate Eq. \eqref{eq:theta_dot_action}, we define a new variable $z=\tan\theta=Q/P$, which simplifies Eq. \eqref{eq:theta_dot_action} 
\begin{align}
\dot{z}=-(1-\mu)+(1+\mu)z^2.
\label{eq:z_dot_final}
\end{align}
Eq. \eqref{eq:z_dot_final}  can be immediately integrated
\begin{align}
z(\tau)=-a\coth\left(\beta\tau\right),
\label{eq:z(t)}
\end{align}
where $a=\sqrt{(1-\mu)/(1+\mu)}$, and $\beta=\sqrt{(1-\mu)(1+\mu)}$.
Therefore, by using the relations
$I=\mu-\cos2\theta,$ and $\cos2\theta=(1-z^2)/(1+z^2),$ we find that
$I(\tau)=(1-\mu^2)/(\cosh(2\beta\tau)-\mu)$, and
\begin{align*}
Q_{\rm sep}(\tau)=
\frac{(1-\mu)\sqrt{2(1+\mu)}\,\cosh(\beta\tau)}
{\cosh(2\beta\tau)-\mu},~
P_{\rm sep}(\tau)=
-\frac{(1+\mu)\sqrt{2(1-\mu)}\,\sinh(\beta\tau)}
{\cosh(2\beta\tau)-\mu}.
\end{align*}

\section{ Energy balance and Melnikov analysis}
\label{sec:melnikov}
As described in the main text, for the non-conservative dynamics, $\mathcal{H}$ is not a conserved quantity, i.e.,  
\begin{align}
\mathcal{\dot H}&=(\partial_Q\mathcal{H})\dot Q+(\partial_P\mathcal{H})\dot P=\partial_Q\mathcal{H}\left[\partial_P \mathcal{H}-\Gamma_l Q-F\sin(\Delta_\omega\tau)\right]+\partial_P\mathcal{H}\left[-\partial_Q \mathcal{H}-\Gamma_lP-F\cos(\Delta_\omega\tau)\right]\nonumber\\
&=-\partial_Q\mathcal{H}\left[\Gamma_l Q+F\sin(\Delta_\omega\tau)\right]-\partial_P\mathcal{H}\left[\Gamma_lP+F\cos(\Delta_\omega\tau)\right]=-(\partial_Q\mathcal{H})\varepsilon_Q-(\partial_P\mathcal{H})\varepsilon_P\ne0.\label{eq:Hdot}
\end{align}
By rearranging Eqs. \eqref{eq:Q_rotating_frame}-\eqref{eq:P_rotating_frame}, we find that $\partial_P\mathcal{H}=\dot Q+\varepsilon_Q$, and $\partial_Q\mathcal{H}=-(\dot P+\varepsilon_P)$. Hence, by substituting these relations into Eq. \eqref{eq:Hdot}, we find that   
\begin{align}
\mathcal{\dot H}=\dot P\varepsilon_Q-\dot Q\varepsilon_P=F[\dot P\sin(\Delta_\omega\tau)-\dot Q \cos(\Delta_\omega\tau)]-\Gamma[P\dot Q-Q\dot P].\label{eq:HdotSim}
\end{align}
On a closed trajectory of the phase portrait $\Delta\mathcal{H}=0$, and therefore,
\begin{align}
\Delta\mathcal{H}\approx\int_0^{\tau_p}\frac{d\mathcal{H}}{d\tau}d\tau=0.
\label{eq:AveH}
\end{align} 
Substituting Eq. \eqref{eq:HdotSim} into Eq. \eqref{eq:AveH}, we find that
\begin{align}
F\approx\Gamma\frac{\int_0^{\tau_p}[P\dot Q-Q\dot P]d\tau}{\int_0^{\tau_p}[\dot P\sin(\Delta_\omega\tau)-\dot Q\cos(\Delta_\omega\tau)]d\tau}. \label{eq:F_per}
\end{align}

\subsection{Melnikov function along the separatrix}
Applying the condition of Eq. \eqref{eq:AveH} on the separatrix trajectory of the conservative system is equivalent to finding the zero of the Melnikov function \cite{ch7}
\begin{align}
M(\tau_0)
={}&
-\int_{-\infty}^{\infty}
\Gamma
\left[Q_{\rm sep}
\frac{\partial \mathcal{H}}{\partial Q}(Q_{\rm sep},P_{\rm sep})
+P_{\rm sep}\frac{\partial \mathcal{H}}{\partial P}(Q_{\rm sep},P_{\rm sep}) \right]d\tau
\nonumber\\
&-
F\int_{-\infty}^{\infty}
\left[
\frac{\partial \mathcal{H}}{\partial Q}(Q_{\rm sep},P_{\rm sep})
\sin\left(\Delta_\omega(\tau+\tau_0)\right)
+
\frac{\partial \mathcal{H}}{\partial P}(Q_{\rm sep},P_{\rm sep})
\cos\left(\Delta_\omega(\tau+\tau_0)\right)
\right]d\tau =0.
\end{align}
The negative work of the damping over the separatrix trajectory can be formally written as
\begin{align}
M_d=-\Gamma I_1(\mu),
\end{align}
where
\begin{align}
I_1(\mu)&=\int_{-\infty}^{\infty}\left[Q_{\rm sep}
\frac{\partial \mathcal{H}}{\partial Q}(Q_{\rm sep},P_{\rm sep})
+P_{\rm sep}\frac{\partial \mathcal{H}}{\partial P}(Q_{\rm sep},P_{\rm sep}) \right]\,d\tau.
\end{align}
The positive work of the external force can be written, using trigonometric identities, as
\begin{align}
M_f(\tau_0)=F\left[I_2(\mu,\Delta_\omega)\sin(\Delta_\omega\tau_0)+I_3(\mu,\Delta_\omega)\cos(\Delta_\omega\tau_0)\right],
\end{align}
where
\begin{align}
I_2(\mu,\Delta_\omega)
&=
-\int_{-\infty}^{\infty}
\left[
\frac{\partial \mathcal{H}}{\partial Q}(Q_{\rm sep},P_{\rm sep})
\cos(\Delta_\omega\tau)
-
\frac{\partial \mathcal{H}}{\partial P}(Q_{\rm sep},P_{\rm sep})
\sin(\Delta_\omega\tau)
\right]d\tau,\\
I_3(\mu,\Delta_\omega)
&=
-\int_{-\infty}^{\infty}
\left[
\frac{\partial \mathcal{H}}{\partial Q}(Q_{\rm sep},P_{\rm sep})\sin(\Delta_\omega\tau)
+\frac{\partial \mathcal{H}}{\partial P}(Q_{\rm sep},P_{\rm sep})\cos(\Delta_\omega\tau)\right]d\tau. 
\end{align}
Due to symmetry, $I_3(\mu,\Delta_\omega)=0$.
Therefore, the Melnikov function becomes
\begin{align}
M(\tau_0)=-\Gamma I_1(\mu)+FI_2(\mu,\Delta_\omega)\sin(\Delta_\omega\tau_0).
\end{align}
where 
\begin{align}
I_1(\mu)
=2\left[\sqrt{1-\mu^2}+2\mu\arctan\left(\frac{1+\mu}{\sqrt{1-\mu^2}}\right)\right],~I_2(\mu,\Delta_\omega)
=
\frac{1}{16\sqrt{2}}\left|
\Re
\left\{
\sqrt{1-\mu}\,\mathcal{S}_1(\mu,\Delta_\omega)
+
\sqrt{1+\mu}\,\mathcal{S}_2(\mu,\Delta_\omega)
\right\}
\right|,\label{eq:melinkov_I}
\end{align}

and the various quantities of $I_2(\mu,\Delta_\omega)$ are given by
\begin{align}
\mathcal{S}_1(\mu,\Delta_\omega)
={}&
(\Delta_\omega-3\ii\beta)\psi\left(\frac{11}{8}+\ii\eta\right)
+
(\Delta_\omega+3\ii\beta)\psi\left(\frac{11}{8}-\ii\eta\right)
+
(3\Delta_\omega+3\ii\beta)\psi\left(\frac{9}{8}-\ii\eta\right)
+
(3\Delta_\omega-3\ii\beta)\psi\left(\frac{9}{8}+\ii\eta\right)
\nonumber\\
&-
4\Delta_\omega
\left[
\psi\left(\frac{7}{8}-\ii\eta\right)
+
\psi\left(\frac{7}{8}+\ii\eta\right)
+
\psi\left(\frac{5}{8}-\ii\eta\right)
+
\psi\left(\frac{5}{8}+\ii\eta\right)
\right]
+
(3\Delta_\omega-3\ii\beta)\psi\left(\frac{3}{8}-\ii\eta\right)
\nonumber\\
&+
(3\Delta_\omega+3\ii\beta)\psi\left(\frac{3}{8}+\ii\eta\right)+
(\Delta_\omega+3\ii\beta)\psi\left(\frac{1}{8}+\ii\eta\right)
+
(\Delta_\omega-3\ii\beta)\psi\left(\frac{1}{8}-\ii\eta\right),\nonumber\\
\mathcal{S}_2(\mu,\Delta_\omega)
={}&
32\beta
+
(\ii\Delta_\omega+3\beta)\psi\left(\frac{11}{8}+\ii\eta\right)
+
(-\ii\Delta_\omega+3\beta)\psi\left(\frac{11}{8}-\ii\eta\right)
+
(3\ii\Delta_\omega-3\beta)\psi\left(\frac{9}{8}-\ii\eta\right)
\nonumber\\
&-
(3\ii\Delta_\omega+3\beta)\psi\left(\frac{9}{8}+\ii\eta\right)
+
4\ii\Delta_\omega
\left[
\psi\left(\frac{7}{8}-\ii\eta\right)
-
\psi\left(\frac{7}{8}+\ii\eta\right)
-
\psi\left(\frac{5}{8}-\ii\eta\right)
+
\psi\left(\frac{5}{8}+\ii\eta\right)
\right]
\nonumber\\
&
-
(3\ii\Delta_\omega+3\beta)\psi\left(\frac{3}{8}-\ii\eta\right)
+
(3\ii\Delta_\omega-3\beta)\psi\left(\frac{3}{8}+\ii\eta\right)+
(-\ii\Delta_\omega+3\beta)\psi\left(\frac{1}{8}+\ii\eta\right)
+
(\ii\Delta_\omega+3\beta)\psi\left(\frac{1}{8}-\ii\eta\right),\nonumber\\
\beta=\sqrt{1-\mu^2},&~~\eta=\frac{\Delta_\omega}{8\beta},~~\psi(z)
=\frac{d\Gamma}{dz}\frac{1}{\Gamma(z)},\nonumber
\end{align}
with $\Gamma(z)$ being the Euler gamma function. Substituting $\mu=0$ into Eq. \eqref{eq:melinkov_I}, simplifies the expressions significantly  and yields
\begin{align} 
F_{\rm cr}=\frac{\Gamma I_1(0)}{I_2(0,\Delta_\omega)}=\frac{2\Gamma}{\pi\Delta_\omega}\ee^{-\pi\Delta_\omega/4}\cosh\left(\frac{\pi\Delta_\omega}{2}\right).
\label{eq:Fcr}
\end{align}

\section{ Floquet analysis, period-doubling instability, and the route to local chaotic attractor}
\label{sec:floquet_pd}
To determine the onset of the first period-doubling instability and the period-doubling cascade towards chaos, we performed a Floquet stability analysis of the periodic responses in the rotating-frame equations. The procedure consists of four steps: (i) computation of the period-one orbit using a shooting method, (ii) integration of the linearized variational equations along this orbit to construct the monodromy matrix, (iii) evaluation of its Floquet multipliers as the direct-drive parameters are varied, and (iv) repeat the same process after each period-doubling instability for the new period-$2^n$ orbit to estimate the Feigenbaum constant.
\\\textbf{\textit{
(i) Periodic-orbit computation:}}
The rotating-frame dynamics can be written in the general form
\begin{align}
\dot{\mathbf{x}}=\mathbf{f}(\mathbf{x},\tau;F,\Delta_\omega),\qquad
\mathbf{x}=
\begin{pmatrix}
Q\\
P
\end{pmatrix},
\label{eq:floquet_system}
\end{align}
where the explicit time dependence originates from the direct drive. For a drive detuning $\Delta_\omega$, the corresponding forcing period in the rotating frame is
\begin{align}
T_d=\frac{2\pi}{\Delta_\omega}.
\label{eq:forcing_period}
\end{align}
A period-one solution satisfies
\begin{align}
\mathbf{x}(T_d;\mathbf{x}_0)=\mathbf{x}_0,
\label{eq:periodic_condition}
\end{align}
where $\mathbf{x}_0=(Q_0,P_0)^T$ is the initial condition. We therefore define the residual
\begin{align}
\mathbf{R}(\mathbf{x}_0)=\mathbf{x}(T_d;\mathbf{x}_0)-\mathbf{x}_0.
\label{eq:shooting_residual}
\end{align}
The periodic orbit is obtained by solving
\begin{align}
\mathbf{R}(\mathbf{x}_0)=\mathbf{0}.
\end{align}
Numerically, the equations of motion were integrated over one forcing period, and the initial condition was iteratively corrected until the final state coincided with the initial state. The periodic branch was followed by numerical continuation in the direct-drive parameters. The converged periodic solution at one parameter value was used as the initial guess for the neighboring value, allowing the period-one solution to be tracked continuously up to its loss of stability.
\\\textbf{\textit{(ii) Variational equations and monodromy matrix:}}
To determine the linear stability of the periodic orbit, we superimpose a small perturbation to the period-one orbit of step (i), $Q(\tau)+\delta Q(\tau),~P(\tau)+\delta P(\tau)$, and linearize the rotating-frame equation with respect to this perturbation. This procedure gives the variational equations
\begin{align}
\begin{pmatrix}
\delta \dot Q\\
\delta \dot P
\end{pmatrix}
=
\begin{pmatrix}
-\Gamma+2QP &
1-\mu+Q^2+3P^2
\\
1+\mu-3Q^2-P^2 &
-\Gamma-2QP
\end{pmatrix}
\begin{pmatrix}
\delta Q\\
\delta P
\end{pmatrix}.
\label{eq:variational_pd}
\end{align}
The coefficients in Eq.~\eqref{eq:variational_pd} are evaluated along the numerically obtained periodic orbit $[Q(\tau),P(\tau)]$.
Defining the fundamental matrix $\mathbf{\Phi}(\tau)$ through
\begin{align}
    \dot{\mathbf{\Phi}}
    =
    \mathbf{J}(\tau)\mathbf{\Phi},
    \qquad
    \mathbf{\Phi}(0)=\mathbf{I},
    \label{eq:fundamental_matrix}
\end{align}
where
\begin{align}
\mathbf{J}(\tau)
=
\begin{pmatrix}
-\Gamma+2QP &
1-\mu+Q^2+3P^2
\\
1+\mu-3Q^2-P^2 &
-\Gamma-2QP
\end{pmatrix}.
\end{align}
The evolution of an arbitrary infinitesimal perturbation is given by
\begin{align}
\delta\mathbf{x}(\tau)=\mathbf{\Phi}(\tau)\delta\mathbf{x}(0).
\end{align}
The rotating-frame equations and the variational equations were integrated simultaneously over one forcing period. The resulting fundamental matrix after one period,
\begin{align}
\mathbf{M}=\mathbf{\Phi}(T_d),
\label{eq:monodromy_matrix}
\end{align}
is the monodromy matrix. It maps an infinitesimal perturbation at the beginning of the period onto the perturbation after one complete cycle,
\begin{align}
\delta\mathbf{x}(T_d)=\mathbf{M}\delta\mathbf{x}(0).
\end{align}
\\ \textbf{\textit{(iii) Floquet multipliers and period-doubling threshold:}}
The eigenvalues of the monodromy matrix,
\begin{align}
\lambda_{\mathrm{Floquet}_{1,2}}=\operatorname{eig}(\mathbf{M}),
\end{align}
are the Floquet multipliers of the period-one orbit. The periodic solution is linearly stable when both multipliers lie inside the unit circle,
\begin{align}
|\lambda_{\mathrm{Floquet}_{1,2}}|<1.
\end{align}
A period-doubling occurs when a real Floquet multiplier ($\lambda_{\mathrm{Floquet}}$) crosses the unit circle through
\begin{align}
\lambda_{\mathrm{Floquet}}=-1.
\label{eq:pd_condition}
\end{align}
At this point, an infinitesimal perturbation changes sign after one forcing period and reproduces itself only after two forcing periods. The original period-one orbit therefore loses stability and a period-two orbit emerges. We calculated the Floquet multipliers while varying the direct-drive parameters and found that one multiplier crosses $\lambda_{\mathrm{Floquet}}=-1$ at
\begin{align}
F_{\rm pd} &= 0.248,\\
\Delta_{\omega_{\rm pd}} &= 2.014.
\label{eq:pd_threshold}
\end{align}
These values therefore define the first period-doubling instability of the period-one response. Below this threshold, the period-one orbit remains stable, whereas beyond it the period-one solution becomes unstable and the system enters the first period-doubled branch.\\
\textbf{\textit{(iv) Feigenbaum constant:}}
To characterize the subsequent route to chaos, we extended the Floquet analysis to higher-order periodic orbits, tracking successive period-doubling bifurcations up to period 64. For each period-$2^n$ orbit, the monodromy matrix was computed over $2^n T_d$, and the corresponding instability threshold $F_n$ was determined by the crossing of a Floquet multiplier through $\lambda_{\mathrm{Floquet}}=-1$. The successive instability thresholds were then used (with $F_{\rm pd}=F_1$) to calculate the finite Feigenbaum ratios,
\begin{align}
\delta_n = \frac{F_{n+1}-F_n}{F_{n+2}-F_{n+1}},
\end{align}
yielding $\delta_1=11.851$, $\delta_2=3.714$, $\delta_3=4.458$, and $\delta_4=4.618$. Although the initial ratios differ considerably from the universal Feigenbaum constant $\delta=4.6692$, the higher-order ratios approach this value, providing numerical evidence consistent with Feigenbaum scaling in the period-doubling cascade to chaos \cite{feigenbaum1978quantitative}.
\section{ Numerical methods and experimental data processing}
\label{sec:numerical_exp}

\subsection{Numerical integration and Poincar\'e sections}
The rotating-frame equations of motion were integrated numerically in MATLAB using the adaptive Runge--Kutta solver \texttt{ode45} with tolerance $\mathrm{RelTol}=\mathrm{AbsTol}=10^{-8}$. For a direct-drive detuning $\Delta_\omega$, the forcing period in dimensionless slow time is

\begin{align}
T_d=\frac{2\pi}{\Delta_\omega}.
\end{align}
For trajectory calculations, the numerical output was evaluated on a uniform grid containing 500 points per forcing period,
\begin{align}
\Delta\tau=\frac{T_d}{500},
\end{align}
while the internal time steps of the adaptive solver were determined automatically.
When direct comparison with an experimental trajectory was required, the numerical initial condition was obtained from the measured quadratures at the beginning of the selected experimental interval. The measured values were linearly interpolated to the exact initial time, rotated into the $(Q,P)$ coordinate system, and converted to dimensionless coordinates using the experimentally determined amplitude conversion factor. In this way, the numerical and experimental trajectories were initialized from the same point in phase space.
Poincar\'e sections were constructed by stroboscopically sampling the trajectory once per forcing period. To suppress dependence on the initial transient, the first $60\%$ of the selected trajectory was discarded. For the numerical trajectory, sampling was performed at
\begin{align}
\tau_n=\tau_{\mathrm{ref}}+nT_d,
\end{align}
where $\tau_{\mathrm{ref}}$ fixes the phase of the direct drive. The values of $Q$ and $P$ at these exact sampling times were obtained by linear interpolation of the numerically integrated trajectory.
The experimental Poincar\'e sections were constructed analogously. For an experimental sampling frequency $f_d$, the stroboscopic period is
\begin{align}
T_{d,\mathrm{exp}}=\frac{1}{f_d}.
\end{align}
The sampling phase was chosen relative to the measured drive according to
\begin{align}
t_{\mathrm{ref}}=\frac{\phi_{\mathrm{exp}}}{2\pi} T_{d,\mathrm{exp}},
\end{align}
and the experimental quadratures were interpolated at
\begin{align}
t_n=t_{\mathrm{ref}}+nT_{d,\mathrm{exp}}.
\end{align}
The phase $\phi_{\mathrm{exp}}$ was selected consistently for each measurement such that the experimental and numerical Poincar\'e sections were compared at the same phase of the periodic forcing.
\subsection{Experimental quadrature processing and phase-portrait alignment}
The experimentally measured in-phase and quadrature signals, denoted here by $X(t)$ and $Y(t)$, were recorded separately and subsequently combined to reconstruct the phase-portrait trajectory. 
The measured quadratures were then rotated by an angle
\begin{align}
    \theta_{\mathrm{exp}}=1.25\pi
\end{align}
according to
\begin{align}
Q_{\mathrm{exp}}
&=X\cos\theta_{\mathrm{exp}}+Y\sin\theta_{\mathrm{exp}},\\
P_{\mathrm{exp}}
&=-X\sin\theta_{\mathrm{exp}}+Y\cos\theta_{\mathrm{exp}}.
\label{eq:experimental_rotation}
\end{align}
This rotation aligns the experimental quadratures with the theoretical rotating-frame coordinates. The conversion between the dimensionless theoretical coordinates and the measured voltage quadratures was determined using the amplitude scale
\begin{align}
\ell=4.970\times10^{-4}\ {\rm V},
\end{align}
such that
\begin{align}
Q_{\mathrm{exp}}=\ell Q,\qquad P_{\mathrm{exp}}=\ell P.
\end{align}
Experimental time was converted to the dimensionless slow time used in the theoretical model through
\begin{align}
\tau=t\lambda/4\omega_1=624t.
\end{align}
For comparisons involving closed phase-space trajectories, a residual translation of the experimental coordinates was applied when required to align the measured stable fixed point with the corresponding theoretical fixed point. The same translation was applied to trajectories belonging to the same well. Any additional rotation or translation used solely for visualization was applied consistently to the complete trajectory and did not alter its geometry.
The lock-in amplifier sampling rate was $13.4\mathrm{kHz}$, and the LPF was of order $4$. 

\subsection{Theoretical conservative and experimental quasi-conservative trajectories}
\begin{figure}[t]
\centering
\includegraphics[width=0.7\linewidth]{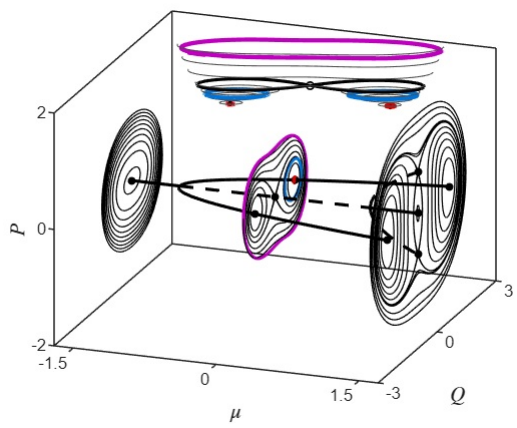}
\caption{The $Q-P$ dynamics as a function of the Hamiltonian single parameter $\mu$. As the parameter $\mu$ increases, the conservative $Q-P$ trajectories change from ($\mu<-1$), solely periodic motions around the origin, to ($-1<\mu<1$) periodic motion around one of the inner wells of a double-well potential or symmetric motion around both wells for larger Hamiltonian values, and finally ($\mu>1$) to motion around inner wells or symmetric motion around the origin or symmetric motion around the three extrema of $\mathcal{H}$. A pair of pitchfork bifurcations (supercritical at $\mu=-1$, and subcritical at $\mu=1$) creates a backbone structure of equilibrium points that is responsible for these qualitative changes of the conservative $Q-P$ trajectories. The solid black backbone branches indicate stable equilibrium points, whereas dashed branches indicate unstable equilibrium points. The thin black contours are graphical representations of the relation $\mathcal{H}(Q,P)=\mathcal{H}_i$ for various Hamiltonian levels $\mathcal{H}_i$, while the additional trajectories at $\mu=0$ are obtained from experimental measurements of periodic orbit when $\varepsilon_{Q,P}\ne0$ -- the red trajectory corresponds to drive values of $\Delta_\omega=10, F=0.145$, in the blue trajectory, $\Delta_\omega=2, F=0.218$, and in the magenta trajectory $\Delta_\omega=1.5, F=0.581$. The inset shows a three-dimensional construction of the double-well potential for $\mu=0$, with the left well showing the analytical trajectories and the right well showing the experimental trajectories.}
    \label{fig:hamiltonian_3d_backbone}
\end{figure}
\begin{figure}[htbp]
    \centering
            \includegraphics[width=3.0in]{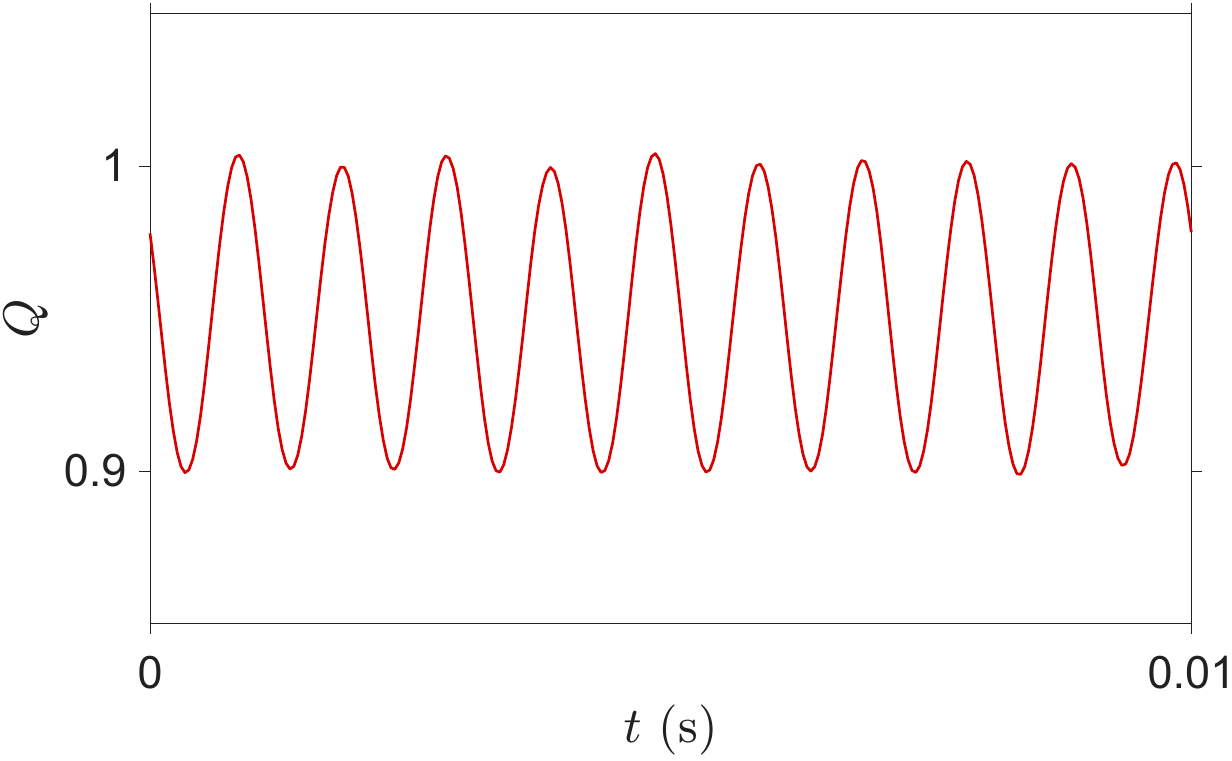}
            \includegraphics[width=3.0in]{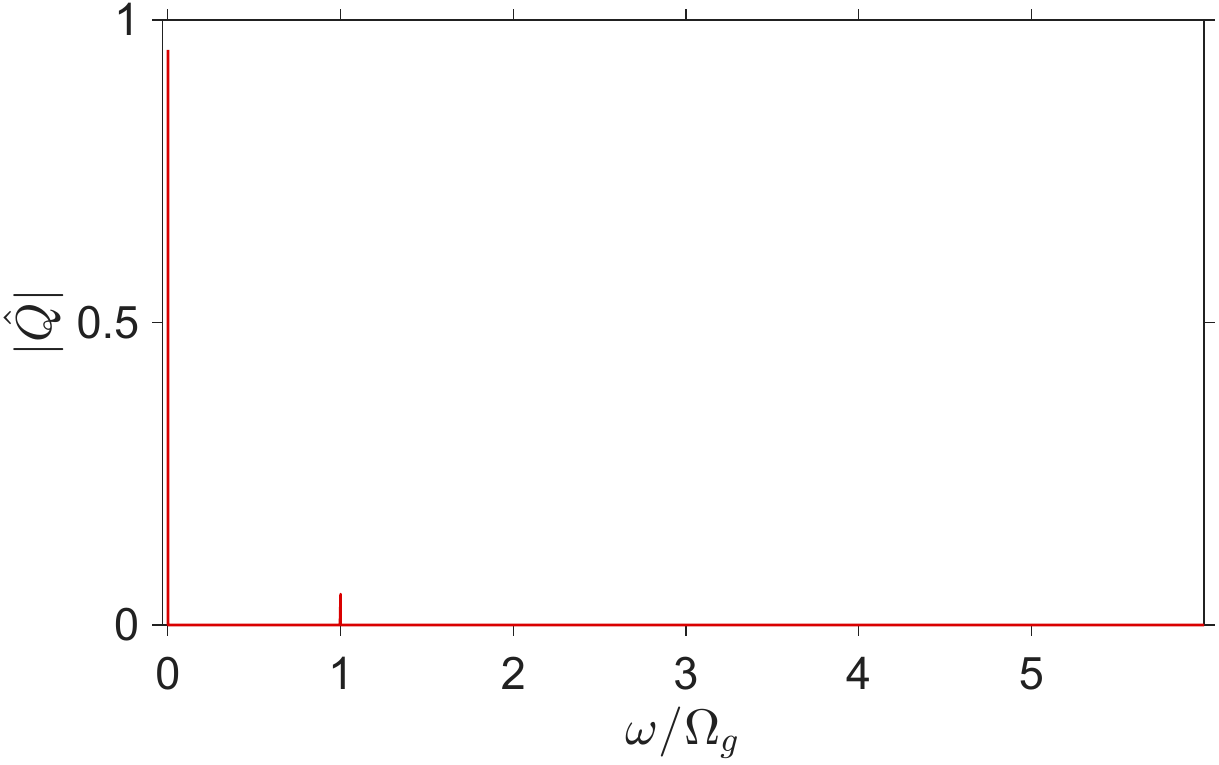}\\
            \includegraphics[width=3.0in]{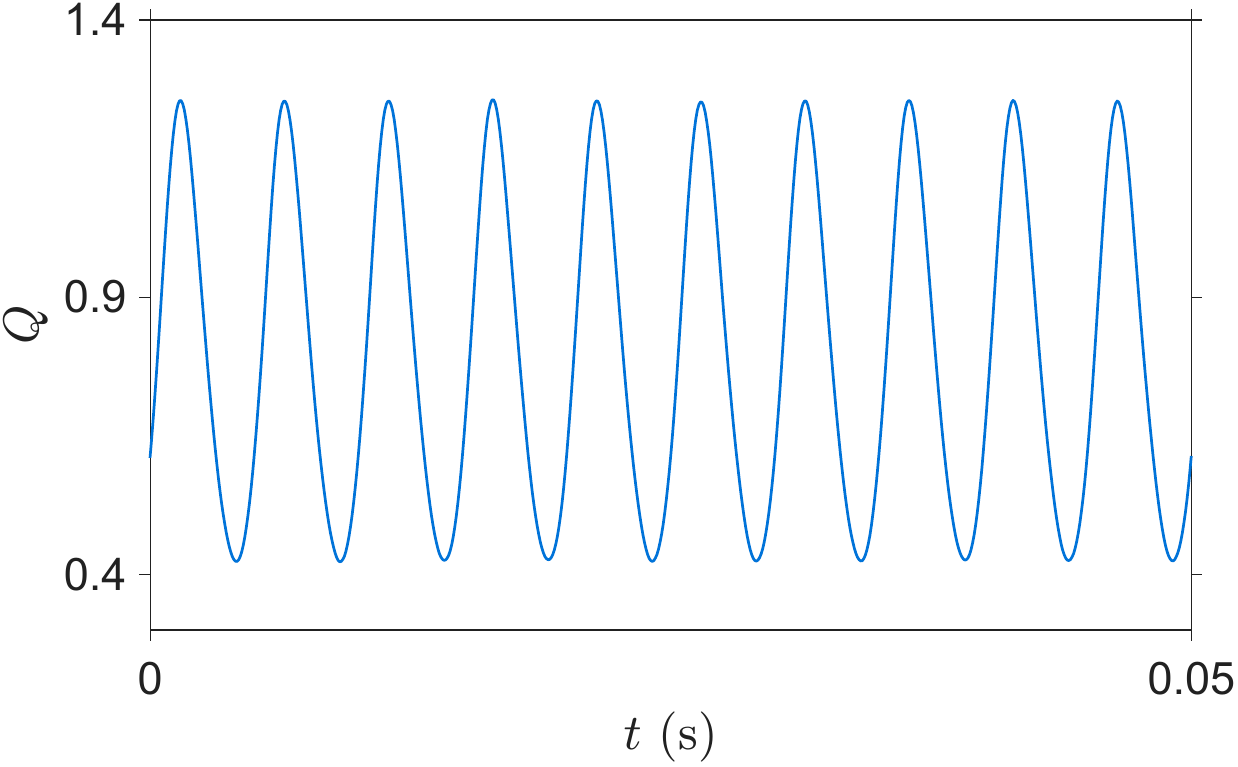}
            \includegraphics[width=3.0in]{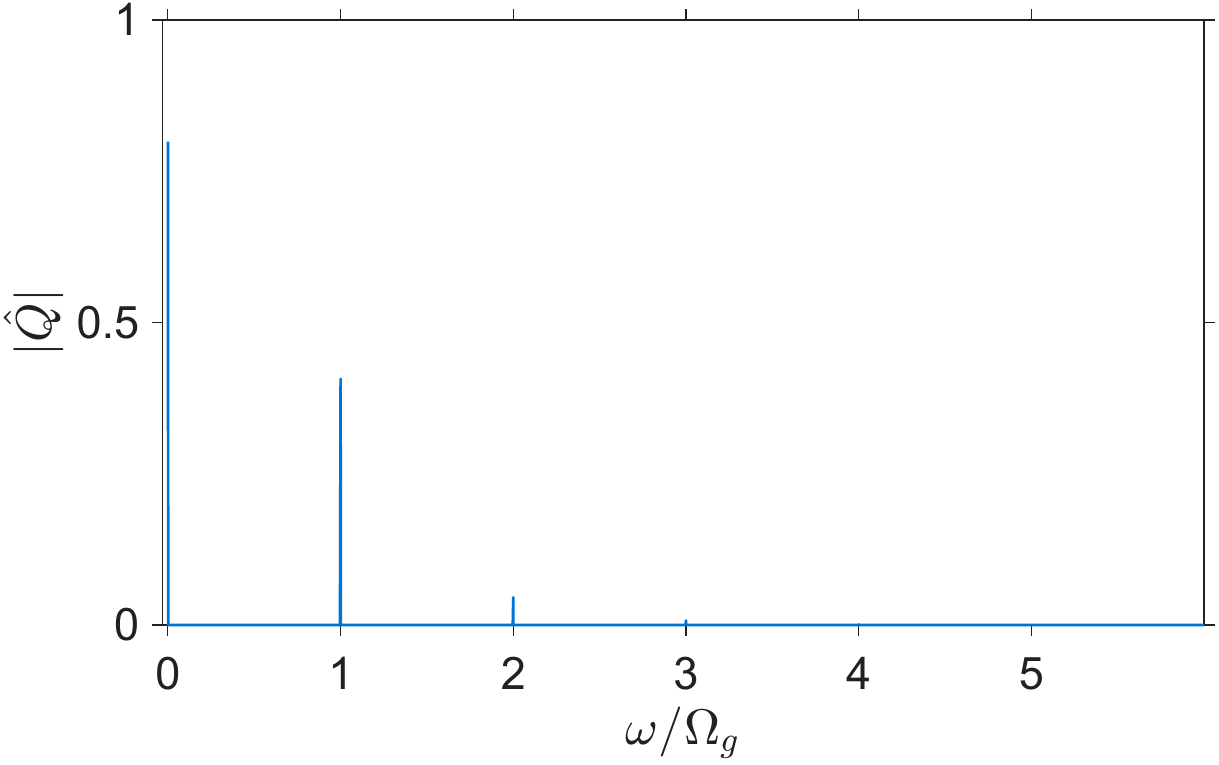}\\
            \includegraphics[width=3.0in]{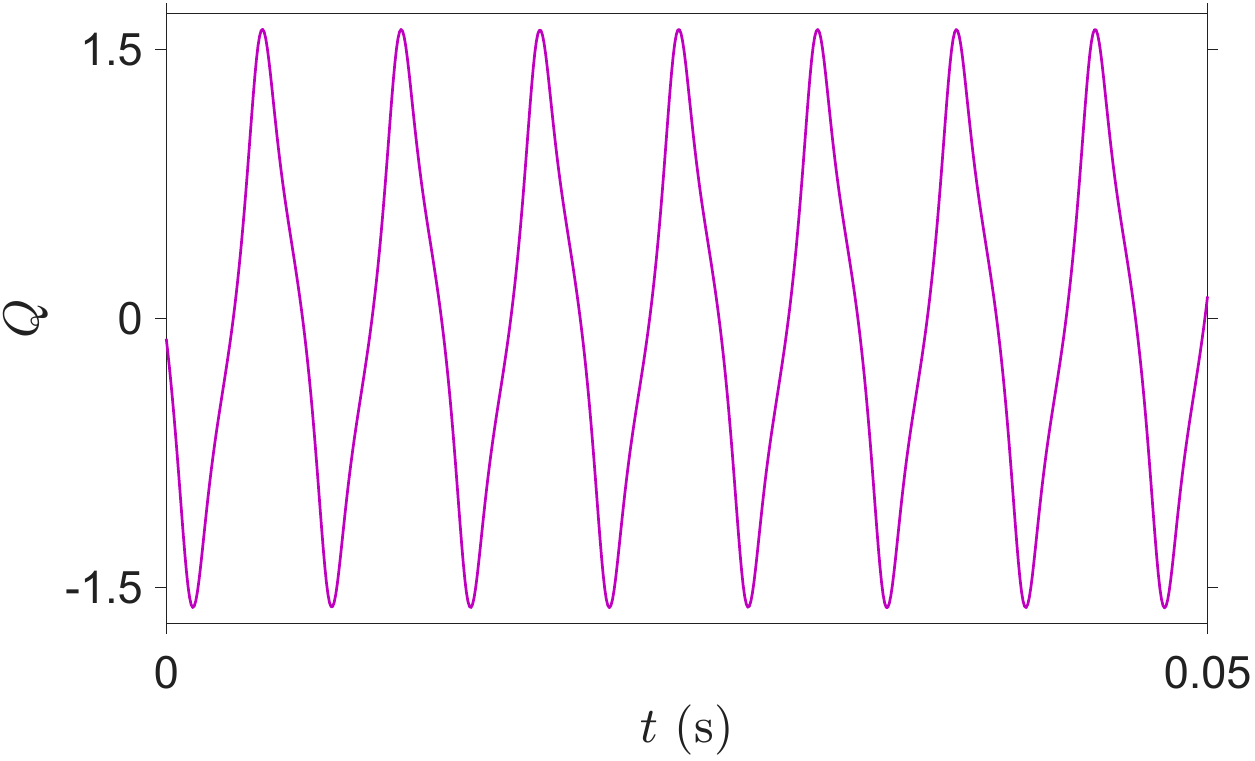}
            \includegraphics[width=3.0in]{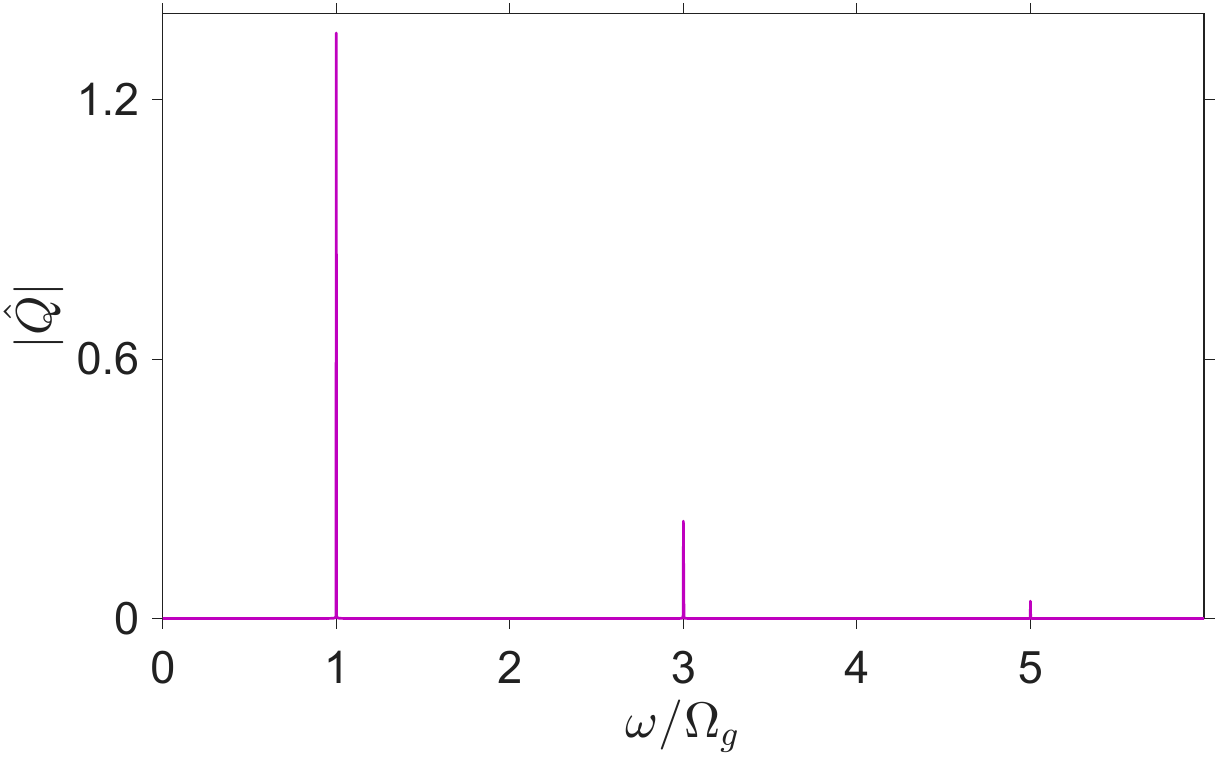}
\caption{Time and frequency domains representation of $Q$ for $\mu=0$ and various Hamiltonian values $\mathcal{H}_i$. Top panel: For $\mathcal{H}_1=-0.24$ slightly above $\mathcal{H}(Q_{\rm min},0)=-0.25$, the $Q$ dynamics are essentially linear and contain a single harmonic, with a frequency $\omega=\Omega_{\rm lin}$, i.e., the constant of the normalized frequency in the Fourier spectrum equals one, $C=\Omega_g/\Omega_{\rm lin}=1$. Center panel: For $0>\mathcal{H}_2=-0.03>\mathcal{H}_1$, the $Q$ dynamics become nonlinear and contain a fundamental harmonic and its even and odd overtones. The softening nonlinearity reduces the oscillation frequency, $C=\Omega_g/\Omega_{\rm lin}=0.63$. Bottom panel: For $0=\mathcal{H}(0,0)<\mathcal{H}_3=0.05$, the $Q$ dynamics are associated with symmetric nonlinear motion around both wells, and contain a fundamental harmonic and its odd overtones. Due to the vicinity of this motion to the separatrix $\mathcal{H}(0,0)$, at which $\omega\rightarrow0$, the fundamental frequency is significantly smaller than $\Omega_{\rm lin}$, $C=\Omega_g/\Omega_{\rm lin}=0.37$. The experimental results correspond to: $\Delta_\omega=10, F=0.145$ for the red top panel, $\Delta_\omega=2, F=0.218$ for the blue center panel, and $\Delta_\omega=1.5, F=0.581$ for the magenta bottom panel.}
    \label{fig:Q_and_FFT_3x2}
\end{figure}

All possible trajectories of the conservative dynamics in the $Q-P$ plane can be viewed as the curves that stem from intersections of the three-dimensional surface $\mathcal{H}(Q,P;\mu)$ with horizontal planes (parallel to the $Q-P$ plane) at variable heights $\mathcal{H}_i=\rm const.$ (Fig. \ref{fig:hamiltonian_3d_backbone}, inset), and therefore, they preserve the inversion $(P\mapsto-P,Q\mapsto-Q)$ symmetry of $\mathcal{H}(Q,P;\mu)$. Moreover, as the parameter $\mu$ varies, two pitchfork bifurcations occur (supercritical at $\mu=-1$, and subcritical $\mu=1$; Fig. \ref{fig:hamiltonian_3d_backbone}), which change the geometry of $\mathcal{H}(Q,P;\mu)$, and therefore also the trajectories in the $Q-P$ plane. Specifically, for $\mu<-1$, $\mathcal{H}(Q,P;\mu)$ has a single minimum, which is the trivial solution $Q=P=\mathcal{H}(0,0;\mu)=0$, and all trajectories correspond to positive Hamiltonian values $\mathcal{H}_i>0$. For $-1<\mu<1$, $\mathcal{H}(Q,P;\mu)$ has two minima, which are given by $Q=\pm Q_{\rm min},~P=0,~\mathcal{H}(\pm Q_{\rm min},0;\mu)=-(\mu+1)^2/4$, where $Q_{\rm min}=\sqrt{\mu+1}$, and the trivial solution $Q=P=\mathcal{H}(0,0;\mu)=0$ turns into a saddle point.  For $\mu>1$, $\mathcal{H}(Q,P;\mu)$ has additional two saddle points at $Q=0,~P=\pm P_{\rm saddle},~\mathcal{H}(0,\pm P_{\rm saddle};\mu)=-(\mu-1)^2/4$, where $P_{\rm saddle}=\sqrt{\mu-1}$, and the trivial solution $Q=P=0$ turns into a local maximum (Fig. \ref{fig:hamiltonian_3d_backbone}). We are mainly interested in the dynamics of $Q$ and $P$ for $-1<\mu<1$, which produce trajectories that are similar to those of the conservative Duffing-Holmes resonator \cite{holmes1979nonlinear}, i.e., motion of a particle in a double-well potential, which is one of the simplest and well-studied models for chaos generation \cite{holmes1979nonlinear,moon1985fractal,wang2023dynamical,cheng2025chaos,rodriguez2026order}.
To compare the experimentally measured motion with the conservative rotating-frame dynamics, we consider the Hamiltonian
\begin{align}
\mathcal{H}(Q,P;0)=\frac{1}{2}(P^2-Q^2)+\frac{1}{4}(Q^2+P^2)^2.
\label{eq:H_num_methods}
\end{align}
For a specified value of $\mathcal H$, the corresponding conservative trajectory was calculated from the analytical solution of the Hamiltonian dynamics. For $\mathcal H<0$, the constant-energy contours form separate closed trajectories around each of the two stable fixed points, whereas $\mathcal H=0$ corresponds to the separatrix. For $\mathcal H>0$, a single constant-energy contour surrounds both wells. For the experimental trajectories shown in the main text, the corresponding theoretical conservative contours were determined using the mean Hamiltonian values calculated from the measured quadratures. These values were approximately $\mathcal{H}\simeq-0.247$ and $-0.115$ for the two single-well responses, and $\mathcal{H}\simeq0.458$ for the interwell response.
To quantify the agreement between the experimental quasi-conservative trajectories and the theoretical conservative trajectories, we compared their actions $\mathcal{J}=\oint_\mathcal{H}PdQ$ (the phase-space areas enclosed by their corresponding orbits). The experimental quadratures were first transformed into the normalized rotating-frame coordinates and aligned with the theoretical phase-space structure. For each experimental trajectory, the action was calculated over complete oscillation cycles and averaged. The theoretical action was obtained from the analytical conservative trajectory at the corresponding mean experimental Hamiltonian, $\langle\mathcal{H}\rangle$. The relative action mismatch was defined as
\begin{align}
\delta_\mathcal{J} =
\frac{\left|\mathcal{J}_{\mathrm{exp}}-\mathcal{J}_{\mathrm{theo}}\right|}
{\mathcal{J}_{\mathrm{theo}}}\times100\%,
\label{eq:area_mismatch}
\end{align}
where $\mathcal{J}_{\mathrm{exp}}$ and $\mathcal{J}_{\mathrm{theo}}$ denote the experimental and theoretical actions, respectively. For the two single-well trajectories, the relative action mismatches were $4.02\%$ and $2.39\%$, respectively, while the interwell trajectory exhibited a mismatch of only $0.62\%$. Thus, the maximum relative discrepancy in the enclosed phase-space area remained below $5\%$ for all three trajectories, demonstrating close agreement between the experimental quasi-conservative dynamics and the underlying conservative phase-space structure.\\
Fig. \ref{fig:Q_and_FFT_3x2} shows the experimental trajectory of $Q(t)$ in various driving frequencies and strengths, where the red trajectory corresponds to drive values of $\Delta_\omega=10, F=0.145$, the blue trajectory, $\Delta_\omega=2, F=0.218$, and the magenta trajectory, $\Delta_\omega=1.5, F=0.581$.
The inner well oscillations in the vicinity of $\pm Q_{\rm min}$ are linear (see top panel of Fig. \ref{fig:Q_and_FFT_3x2}), and contain a single harmonic $\Omega_{\rm lin}=2\pi/\tau_p(m_{\mathcal{H}_i}=0)=2\sqrt{1+\mu}$. Closer to the separatrix, the inner well oscillations are nonlinear (see center panel of Fig. \ref{fig:Q_and_FFT_3x2}), and their frequency content is associated with a fundamental frequency and its even and odd overtones. At the separatrix, the oscillation frequency goes to zero, and the quadratures exhibit exponential time dependence. Above the separatrix, where the oscillations of the quadratures are large, symmetric with respect to the origin, around both wells (see bottom panel of Fig. \ref{fig:Q_and_FFT_3x2}), the frequency content is associated with a fundamental frequency and only its odd overtones. 
\FloatBarrier
These trajectories demonstrate how the measured dissipative and driven trajectories relate to the underlying conservative phase-space structure.

\subsection{Lyapunov exponent calculation}

The Lyapunov exponents were calculated numerically by integrating the rotating-frame equations together with their variational equations and periodically orthonormalizing a set of tangent vectors. Because the rotating-frame equations contain an explicit periodic forcing term, the non-autonomous two-dimensional system was written as an equivalent autonomous three-dimensional system by introducing the time variable as an additional state,
\begin{align}
    \mathbf{x}
    =
    \begin{pmatrix}
        Q\\
        P\\
        s
    \end{pmatrix},
    \qquad
    \dot s=1,
\end{align}
where the forcing phase is $\Delta_\omega s$. The dynamical equations used for the calculation were
\begin{align}
\dot Q&=(1-\mu)P+PQ^2+P^3-\Gamma Q-F\sin(\Delta_\omega s),
\nonumber\\
\dot P&=(1+\mu)Q-QP^2-Q^3-\Gamma P-F\cos(\Delta_\omega s),
\label{eq:lyapunov_system}\\
\dot s &=1.\nonumber
\end{align}
An infinitesimal perturbation $\delta\mathbf{x}$ evolves according to
\begin{align}
    \delta\dot{\mathbf{x}}
    =
    \mathbf J(\mathbf{x})
    \delta\mathbf{x},
\end{align}
where $\mathbf J$ is the Jacobian of Eq.~\eqref{eq:lyapunov_system}. The physical trajectory and three initially orthonormal tangent vectors were integrated simultaneously. After an initial transient of $\tau_{\mathrm{tr}}=3.0\times10^4$ the tangent vectors were evolved over successive intervals $T_{\mathrm{orth}}=0.5.$ After every interval, a QR decomposition was performed,
$\mathbf V=\mathbf Q_R\mathbf R,$ where the columns of $\mathbf V$ are the evolved tangent vectors. The orthonormal matrix $\mathbf Q_R$ was used to initialize the tangent vectors for the next interval, while the logarithms of the diagonal elements of $\mathbf R$ were accumulated. The Lyapunov exponents were then evaluated as
\begin{align}
    \lambda_{{\mathrm{Lyap},i}}
    =
    \frac{1}{KT_{\mathrm{orth}}}
    \sum_{k=1}^{K}
    \ln\left|R_{ii}^{(k)}\right|,
    \label{eq:lyapunov_qr}
\end{align}
where $K=4\times10^4$ orthonormalization intervals were used. The total accumulation time following transient removal was therefor $T_{\mathrm{L}}=KT_{\mathrm{orth}}=2.0\times10^4.$
The equations of motion and variational equations were integrated using \texttt{ode45} with relative and absolute tolerances of $10^{-8}$.
A positive Lyapunov exponent, $\lambda_{\mathrm{Lyap}}>0$, indicates exponential sensitivity to initial conditions and was therefore used as an additional numerical signature of chaotic dynamics. One exponent of the autonomous extended system is expected to approach zero, corresponding to perturbations along the time/forcing-phase direction, and the sum of the other two exponents is negative and equal to the negative magnitude of the decay rate $-2\Gamma$ \cite{sandri1996numerical}.

\subsection{Global transport, stretching, and folding}
\label{sec:stretching_folding}
\begin{figure}[htbp]
    \centering
    \includegraphics[width=\linewidth]{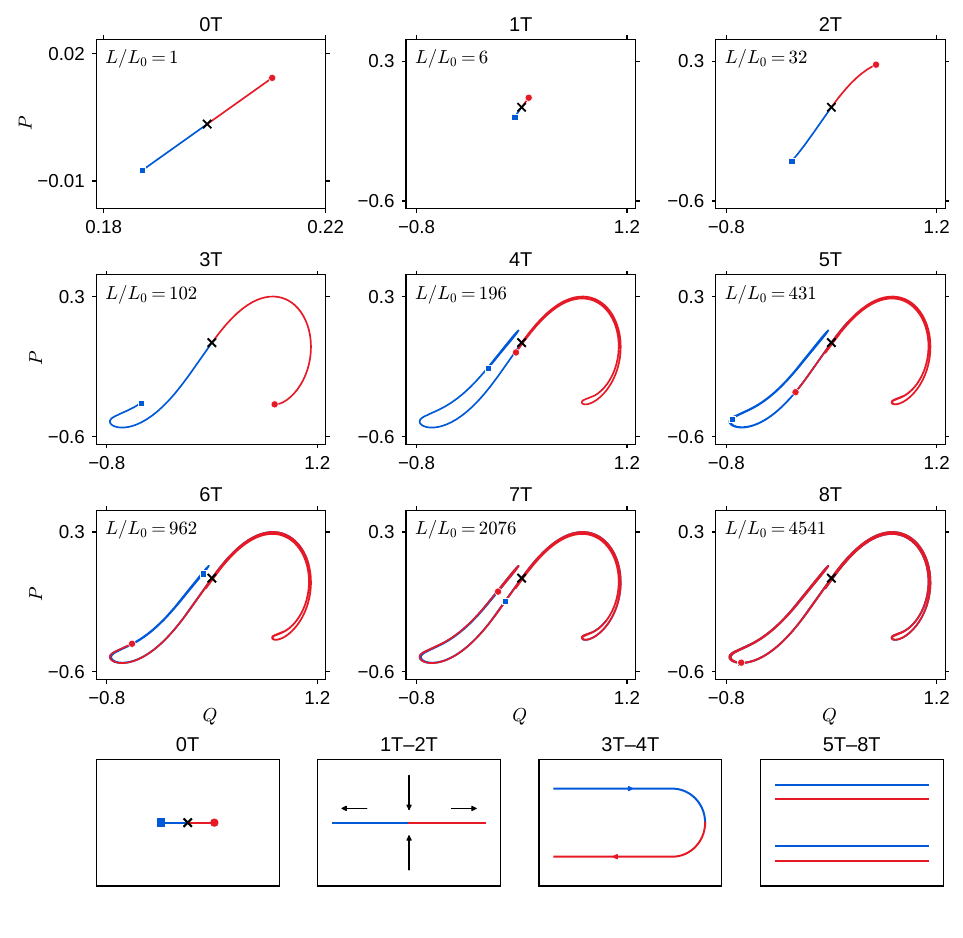}
    \caption{
    Stretching and folding in the rotating-frame phase space for $F=0.339$, $\Delta\omega=2.014$, $\Gamma=0.428$, and $\mu=0$. The upper nine panels show successive stroboscopic images of a short initial segment aligned with the local unstable direction of the saddle periodic orbit, sampled at intervals $T=2\pi/\Delta\omega$. Blue and red identify the two halves of the initial segment, with squares and circles marking their respective endpoints. Black crosses indicate the saddle fixed point of the stroboscopic map, and $L/L_0$ gives the segment length relative to its initial value. The $0T$ panel is magnified. Stretching near the saddle and subsequent excursions around the wells produce folds and closely spaced strands. The bottom row schematically illustrates the corresponding stages of the idealized horseshoe map.
    }
    \label{fig:stretching_folding}
\end{figure}

The geometric mechanism underlying cross-well chaos is illustrated in Fig.~\ref{fig:stretching_folding}, which shows successive stroboscopic images of a short segment initially aligned with the unstable direction of the saddle periodic orbit. Sampling at intervals of the drive period $T=2\pi/\Delta\omega$ defines a two-dimensional Poincaré map, in which the saddle periodic orbit appears as a fixed point. Nearby initial conditions separate along its unstable direction and approach one another along its stable direction. Consequently, a small neighborhood is elongated and compressed into a thin strip. The initial segment shown in the figure resolves the elongation, with its relative length increasing from $L/L_0=1$ to approximately $32$ within two drive periods. The blue and red colors retain the identity of the two original halves throughout their subsequent deformation.

As the stretched segment leaves the saddle neighborhood, its two halves make excursions around the two wells. The nonlinear flow bends these excursions back toward the saddle, creating the first visible folds at $3T$–$4T$. Portions of the returning segment then pass close to portions that began elsewhere along it, producing adjacent strands with different histories. Further iterations stretch these returning portions again, generating additional folds, so that by $8T$ both original halves extend around both wells and the total segment length exceeds its initial value by more than three orders of magnitude. The growth of length within a bounded region of phase space produces increasingly fine structure, while contraction transverse to the strands brings them into close proximity.

This combination of stretching, contraction, folding, and return is the geometric mechanism represented by Smale's horseshoe map \cite{Holmes1989}, illustrated schematically in the bottom row. In the idealized construction, an initially compact region is stretched into a thin strip and folded so that distinct portions return across the original region. Repetition creates progressively finer strips, and trajectories can follow different sequences of passages through the returning branches. In the oscillator, the corresponding branches are organized by excursions around the wells and returns near the saddle. The numerical sequence illustrates this horseshoe mechanism, while the transverse manifold intersections predicted by the Melnikov analysis provide its dynamical basis.

For $F>F_{\rm M}$, the stable and unstable manifolds of the saddle periodic orbit intersect transversely and form a homoclinic tangle within the Melnikov approximation. Segments of these manifolds bound lobes that are carried across the boundary inherited from the unperturbed separatrix. Successive lobe passages therefore provide a route for transport between the wells. The figure connects this global transport to the local instability of the saddle by showing how a short initial segment is stretched, returned, and redistributed across both wells. The rate of interwell transport depends on the accessibility and size of the transport lobes, allowing long episodes of motion within either well even when cross-well passages are possible.
\FloatBarrier

\section{ Merging crisis and residence-time scaling}
\label{sec:crisis}
The transition from local single-well chaos to global double-well chaos was characterized by measuring the residence time between successive interwell transitions. Near the merging crisis, chaotic trajectories exhibit long episodes of motion confined to either well, interrupted by transitions between the two wells. The mean residence time is expected to follow the crisis-induced intermittency scaling law \cite{grebogi1983crises}
\begin{align}
\frac{\langle T_{\mathrm{res}}\rangle}{T}=
C(F-F_{\mathrm{cr}})^{-\eta},
\label{eq:crisis_scaling_sm}
\end{align}
where $T$ is the period of the secondary forcing, $F_{\mathrm{cr}}$ is the critical forcing amplitude, $C$ is a proportionality constant, and $\eta$ is the critical scaling exponent.

\subsection{Identification of interwell switching events}
Interwell transitions were identified from stroboscopic samples of the rotating-frame in-phase component $Q$. For the numerical trajectories, the stroboscopic sampling period was
\begin{align}
T=\frac{2\pi}{\Delta_\omega},
\end{align}
where $\Delta_\omega=2.014$ is the dimensionless forcing frequency. For the experimental measurements, the quadratures were sampled at the corresponding forcing period
\begin{align}
T_{\mathrm{exp}}=\frac{1}{f_{\mathrm{P}}}
\simeq 5.0\,\mathrm{ms},
\end{align}
with $f_{\mathrm{P}}=200.03\,\mathrm{Hz}$. A fixed sampling phase was maintained throughout each measurement. The experimental quadratures were rotated by $1.25\pi$ and normalized using the experimentally determined scale $\ell_{\mathrm{exp}}=4.97\times10^{-4}\,\mathrm{V}$, allowing the same dimensionless switching criterion to be applied to the numerical and experimental data.
To identify transitions while avoiding spurious switches caused by fluctuations near the origin, we employed a hysteretic well-assignment criterion. A trajectory was assigned to the right well when $Q>Q_{\mathrm{th}}$, and to the left well when $Q<-Q_{\mathrm{th}}$, with $Q_{\mathrm{th}}=0.4$.
Once assigned to a well, the trajectory retained its current state until the opposite threshold was crossed. Thus, a transition from the right well to the left well was registered only after $Q<-Q_{\mathrm{th}}$, and vice versa. This procedure suppresses repeated switching events associated with fluctuations around $Q=0$. The same threshold was used for both numerical and experimental trajectories.

\subsection{Residence time calculation}
To calculate the residence time, we let $n_i$ denote the stroboscopic index at which the $i_{\rm th}$ interwell transition was detected. The residence time between two consecutive transitions was calculated as
\begin{align}
T_{\mathrm{res},i}=(n_{i+1}-n_i)T.
\end{align}
The normalized residence time is therefore
\begin{align}
\frac{T_{\mathrm{res},i}}{T}
=n_{i+1}-n_i.
\end{align}
For each forcing amplitude, the mean normalized residence time was calculated from the $N$ complete intervals,

\begin{align}
\frac{\langle T_{\mathrm{res}}\rangle}{T}=
\frac{1}{N}
\sum_{i=1}^{N}
(n_{i+1}-n_i).
\label{eq:mean_residence_sm}
\end{align}
Only intervals bounded by two detected switching events were included. The initial interval, preceding the first detected transition, and the final interval, following the last detected transition, were excluded because their full durations were not observed.
The numerical trajectories were obtained by integrating the rotating-frame equations using a fourth-order Runge--Kutta method, with $\Gamma=0.428$, $\mu=0$, and $\Delta_\omega=2.014$.
The secondary forcing phase was fixed at $2\pi/3$. For each forcing amplitude, four different initial conditions were integrated. The first $5000$ forcing periods were discarded as transients, followed by $10000$ retained periods per trajectory. The integration used $128$ time steps per forcing period.
The experimental residence-time statistics were obtained from four recordings at driving voltages of $70.25$, $70.50$, $70.75$, and $71.00\,\mathrm{mV}$. The corresponding numbers of detected transitions were $32$, $107$, $173$, and $211$, respectively. The switching rates increased with the driving amplitude, corresponding to shorter mean residence times.

\subsection{Estimation of the critical forcing amplitude}
The critical forcing amplitudes were treated separately for the numerical and experimental systems. This distinction accounts for possible differences between the numerical dynamics and the experimentally observed transition. For the numerical results, the reference value $F_{\mathrm{cr}}^{\mathrm{num}}=0.333673$
was used to calculate $F-F_{\mathrm{cr}}^{\mathrm{num}}$. This value was specified as a provisional numerical reference in the numerical analysis and was not independently estimated by the residence-time fitting procedure.
For the experimental measurements, the dimensional driving voltage $U_d$ was converted into the dimensionless forcing amplitude using $F=k_F U_d$,
where $U_d$ is expressed in millivolts and
$k_F=0.339045/70\simeq 4.8435\times10^{-3}\,\mathrm{mV}^{-1}$.
The experimental critical voltage was estimated by fitting the residence-time scaling relation with the theoretical exponent fixed at $\eta=0.824$. The fitting procedure minimized the squared residuals in logarithmic space, with the critical voltage constrained to lie below the lowest measured driving voltage. The resulting estimate of the critical voltage was converted into the corresponding dimensionless critical forcing amplitude using
$F_{\mathrm{cr}}^{\mathrm{exp}}=k_F U_{\mathrm{cr}}$.
The experimental and numerical data were then plotted relative to their respective critical forcing amplitudes.

\subsection{Theoretical scaling exponent from Floquet multipliers}
The scaling exponent associated with crisis-induced intermittency can be related to the local stability of a saddle periodic orbit involved in the global bifurcation. To obtain a theoretical prediction for the exponent, we calculate the Floquet multipliers of the central saddle periodic orbit of the rotating-frame equations.
We first identify the periodic orbit by numerical continuation from the equilibrium at $(Q,P)=(0,0)$ for $F=0$. The forcing amplitude is gradually increased, and the corresponding periodic orbit is determined using a Newton shooting method. Denoting the state vector by $\mathbf{z}=(Q,P)^T$ and the flow over one
forcing period by $\boldsymbol{\phi}_T$, the periodicity condition is
\begin{align}
\boldsymbol{\phi}_T(\mathbf{z}_0)
-\mathbf{z}_0=0,
\label{eq:periodic_shooting_crisis}
\end{align}
where $T=2\pi/\Delta_\omega$ is the forcing period. The Newton correction to the initial condition is obtained from
\begin{align}
\left[M(\mathbf{z}_0)-I\right]
\delta\mathbf{z}_0=
-\left[
\boldsymbol{\phi}_T(\mathbf{z}_0)
-\mathbf{z}_0
\right],
\end{align}
where $M$ is the monodromy matrix and $I$ is the identity matrix. The monodromy matrix is determined as described by Eqs. \eqref{eq:fundamental_matrix}-\eqref{eq:monodromy_matrix}. Its eigenvalues are the Floquet multipliers $\beta_u$ and $\beta_s$, corresponding to the expanding and contracting directions of the saddle periodic orbit, respectively,
\begin{align}
|\beta_u|>1,
\qquad
0<|\beta_s|<1.
\end{align}
For the homoclinic crisis-scaling relation, the corresponding candidate critical exponent is given by
\begin{align}
\eta_{\mathrm{hom}}=
\frac{\ln|\beta_s|}
{2\ln|\beta_u\beta_s|}.
\label{eq:gamma_floquet_crisis}
\end{align}
The calculation was performed using
$\mu=0$, $\Gamma=0.428$,
$\Delta_\omega=2.014$, and
$F_{\rm M}=0.3302$.
The calculation yields the homoclinic candidate exponent $\eta_{\mathrm{hom}}\simeq0.824$. The independent power-law fits to the numerical and experimental data, in the main text, yield $\eta_{\mathrm{num}}=0.892$ ($R^2=0.939$) and $\eta_{\mathrm{exp}}=0.815$ ($R^2=0.997$), which are close to the Floquet-based calculation.

\subsection{Comparison of theoretical, numerical and experimental scaling}
To test the predicted scaling, the measured residence times were compared with Eq.~\eqref{eq:crisis_scaling_sm}. For a fixed critical amplitude and exponent, the scaling law can be written in logarithmic form as
\begin{align}
\ln\left(
\frac{\langle T_{\mathrm{res}}\rangle}{T}
\right)
=
\ln C-\eta_{\mathrm{th}}
\ln(F-F_{\mathrm{cr}}).
\label{eq:log_crisis_scaling_sm}
\end{align}
The proportionality coefficient $C$ was determined by minimizing the squared residuals in logarithmic space. Defining
\begin{align}
x_i=\ln(F_i-F_{\mathrm{cr}}),
\qquad
y_i=\ln\left(
\frac{\langle T_{\mathrm{res}}\rangle_i}{T}
\right),
\end{align}
the fitted intercept is given by
\begin{align}
\ln C=
\frac{1}{N}
\sum_{i=1}^{N}
\left(y_i+\eta_{\mathrm{th}}x_i\right).
\end{align}
The agreement between the observed residence times and the predicted scaling was quantified using the coefficient of determination in logarithmic space,
\begin{align}
R^2=
1-
\frac{
\sum_i(y_i-\hat y_i)^2
}{
\sum_i(y_i-\bar y)^2
},
\label{eq:r2_crisis_sm}
\end{align}
where $\hat y_i=\ln C-\eta_{\mathrm{th}}x_i$ and $\bar y$ is the mean of the observed logarithmic residence times.
For the numerical data, the critical forcing amplitude and theoretical exponent were fixed, and only the prefactor $C$ was fitted. For the experimental data, the critical forcing amplitude and prefactor were fitted while maintaining the theoretical exponent $\eta_{\mathrm{th}}=0.824$.
The resulting trends demonstrate a decrease in the mean residence time with increasing distance from the critical forcing amplitude,
consistent with crisis-induced intermittency.

\subsection{Statistical uncertainty and robustness}
The uncertainty in the mean normalized residence time was estimated from the distribution of complete residence intervals at each forcing amplitude. For $N$ complete intervals, the sample standard deviation is

\begin{align}
s=
\sqrt{
\frac{1}{N-1}
\sum_{i=1}^{N}(r_i-\bar r)^2
},
\end{align}
where $r_i=T_{\mathrm{res},i}/T$ and $\bar r$ denotes the sample mean. An approximate $95\%$ confidence interval for the mean residence time was calculated as
\begin{align}
\frac{\langle T_{\mathrm{res}}\rangle}{T}
\pm 1.96\frac{s}{\sqrt{N}}.
\label{eq:residence_ci_sm}
\end{align}
This uncertainty estimate was applied to both numerical and experimental residence-time measurements. The expression assumes that the intervals can be treated as approximately independent. Temporal correlations between consecutive residence intervals may affect the effective sample size and therefore the confidence interval width. Since residence times may exhibit broad distributions near the crisis, the normal approximation should be interpreted as an estimate of sampling uncertainty rather than an exact confidence bound. Furthermore, the numerical and experimental datasets contain different numbers of complete residence intervals, reflecting the different simulation and measurement durations. Near the critical forcing amplitude, the longer residence intervals reduce the number of observed transitions and increase the statistical uncertainty.

\clearpage
\bibliography{Amitbib}